\documentclass[fleqn,10pt]{wlscirep}
\usepackage[utf8]{inputenc}
\usepackage[T1]{fontenc}
\usepackage{amsmath}
\usepackage{amsfonts}
\usepackage{amssymb}
\usepackage{graphicx}
\usepackage{empheq}
\usepackage{bm}

\newcommand{\tr}{\mbox{tr}}

\newcommand{\txtpow}[1]{{\mbox{\scriptsize{#1}}}}

\newcommand{\boldrho}{{\boldsymbol{\rho}}}
\newcommand{\avg}[1]{{\langle #1 \rangle}}

\DeclareSymbolFont{bbold}{U}{bbold}{m}{n}
\DeclareSymbolFontAlphabet{\mathbbold}{bbold}

\title{Field Correlations in Surface Plasmon Speckle}

\author[1,*]{Matthew R. Foreman}
\affil[1]{Blackett Laboratory, Imperial College London, Prince Consort Road, London, SW7 2AZ, United Kingdom }

\affil[*]{matthew.foreman@imperial.ac.uk}

\begin{abstract}
In this work fluctuations in the electric field of surface plasmon polaritons undergoing random scattering on a rough metallic surface are considered. A rigorous closed form analytic expression is derived describing second order correlations in the resulting plasmon speckle pattern assuming statistically stationary and isotropic roughness. Partially coherent planar Schell-model source fields can also be described within the developed framework. Behaviour of the three-dimensional degree of cross polarisation and spectral degree of coherence is also discussed. Expressions derived take full account of dissipation in the metal with non-universal behaviour exhibited within the correlation length of the surface and source fields.
\end{abstract}
\begin{document}

\flushbottom
\maketitle

\thispagestyle{empty}

\section*{Introduction}
Wave propagation in disordered media can give rise to a variety of interesting physical phenomena such as Anderson localisation and coherent back scattering \cite{Sheng1995,vanRossum1998,Berkovitsa1994}. Ultimately such coherent effects derive from phase correlations which persist between elementary scattered waves in spite of the randomisation imparted by the media. Intensity correlations are also known to exist in the random interference pattern, or speckle, produced by wave scattering in such media. For example, short range intensity correlations, often denoted $C_1$, give rise to the characteristic size of an individual speckle \cite{Goodman2006,Dogariu2015}, whereas in the multiple scattering regime longer range $C_2$ and $C_3$ correlations are responsible for enhanced mesoscopic fluctuations \cite{Feng1988} and deviations from standard Rayleigh statistics \cite{Shnerb1991,Kogan1993}. An infinite range space-time correlation, $C_0$, has also been predicted in speckle patterns produced by point sources embedded in disordered media \cite{Skipetrov2000}. 
Developments in the theoretical understanding of such spatial and temporal correlations of light in scattering media \cite{Feng1988,Berkovitsa1994} have in turn enabled a number of novel experimental techniques in, for example, imaging \cite{Bertolotti2012}, particle analysis \cite{Berne2000} and wave control \cite{Lemoult2009,Bender2017}. Recently it has been recognised that evanescent contributions to the field inside a random medium, or at sub-wavelength distances from its exit surface,  can  significantly modify the properties of speckle patterns, by introducing sub-wavelength correlations \cite{Apostol2003,Carminati2010} and a polarisation dependence \cite{Liu2005f,Laverdant2008,Parigi2016a}. Non-universal behaviour \cite{Greffet1995a,Caze2010}, whereby the statistical properties of the scattered field are strongly dependent on the statistics of the surface, can also occur in contrast to far-field speckle \cite{Nussenzveig1987,Ponomarenko2002}. 

Resonant surface waves, such as surface exciton-polaritons \cite{Takatori2017}, surface phonon-polaritons \cite{LeGall1997}, or surface plasmon-polaritons (SPPs) \cite{Barnes2003}, can also play an important role in scattering at interfaces and thus can affect the physical properties of random electromagnetic fields. Spectral variations with distance from a surface can, for instance, arise due to  the characteristic exponential fall off of surface phonon polaritons in silicon carbide \cite{Shchegrov2000}, whereas SPP scattering can give rise to a flat absorption band in the infra-red part of the spectrum when supported in random metallo-dielectric films close to the percolation threshold \cite{Gadenne1989}. Spatial correlations in the near field speckle intensity have also been considered in a number of studies, partly due to their experimental accessibility. Both oscillatory and monotonically decaying intensity correlation functions have for instance been observed near semi-continuous metal interfaces, where the dominant behaviour is dictated by the degree of SPP scattering \cite{Seal2005}. Signatures of random SPP scattering can also be exhibited in far-field intensity correlations \cite{VanBeijnum2012}. Intensity correlations however do not capture lower order statistical properties such as correlations between different field components. Field-field correlations, as parametrised by the cross spectral density matrix \cite{Mandel1995}, have thus also seen attention in the literature and capture both coherence and polarisation properties of randomly scattered fields. Strong near field polarisation effects in the presence of SPPs or surface phonons have for example been predicted \cite{Setala2002a,Carminati1999}, whereas surface waves excited by thermal emission from a half space have been shown to give rise to long coherence lengths  \cite{Henkel2000}. Anisotropy in the coherence length has also been analysed \cite{Caze2013a}. Notably, the cross-spectral density matrix can also be related to the density of states in disordered systems \cite{Carminati2015,Joulain2005a}, which is of fundamental importance since it governs many light matter interactions, such as fluorescence and thermal emission. Fluctuations in the local density of states near rough and fractal metallic films, resulting from random interference of SPPs, have been experimentally observed and linked to SPP localisation \cite{Castanie2012,Krachmalnicoff2010}.

In this work we consider the field correlations present in the random scattering of surface plasmon polaritons (SPPs) propagating on a rough surface in closer detail. 
Specifically, we provide fully analytic closed-form expressions for correlations, present in the roughness induced field fluctuations in SPP speckle (Eqs.~\eqref{eq:WSPPfinal}--\eqref{eq:Lmn}). Our derivation exploits the known analytic form of the Green's tensor describing SPP scattering \cite{Sondergaard2004}.  These formulae help generalise and validate the predications of more specialised and approximate treatments given in the literature, which for example either assume the exciting source or surface correlation function is well described by a Dirac delta function (e.g. by considering a thermal source), or factor out slowly varying terms from the full integral representation of the cross-spectral density matrix \cite{Carminati1999,Henkel2000,Setala2002a}. Our formulae therefore afford greater physical insight by enabling description of more general surface correlation functions where the only restriction made is that of isotropic stationary statistics. Due to the complexity of the formulae we restrict attention to the single scattering regime, such that we can employ the well known equivalent surface current model for scattering from random metal interfaces as initially described by Kr\"oger and Kretschmann \cite{Kroger1970,Bousquet1981}. We conclude by using our formulae to study the spatial polarisation and coherence properties of SPP speckle. The former is possible since our derivation yields the full correlation matrix, including off-diagonal terms. Such properties can be relevant in, for example, plasmonic interferometry \cite{Gao2013,Feng2012a}, plasmonic focusing \cite{Lee2010a} and leakage radiation microscopy \cite{Drezet2008a,Hohenau2011}.

\section*{Theoretical derivations}

\begin{figure}[t!]
	\begin{center}
		\includegraphics[width=0.5\columnwidth]{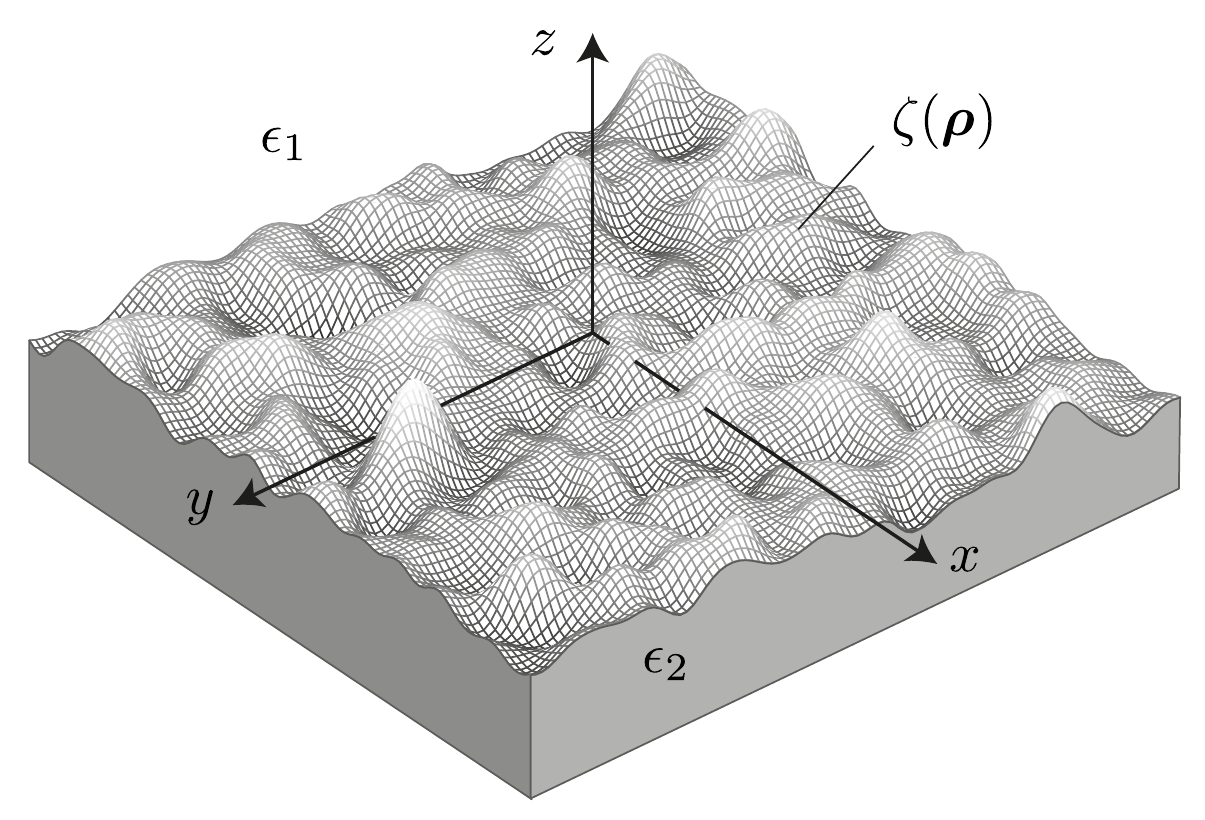}
		\caption{Schematic of a rough surface $\zeta(\boldrho)$ separating a dielectric and metal electric of permittivity $\epsilon_1$ and $\epsilon_2$ respectively. }\label{fig:surface}
	\end{center}
\end{figure}

We consider the system shown in Figure~\ref{fig:surface} which is composed of a dielectric medium of electric permittivity $\epsilon_1$ in the region $z > \zeta(\boldrho)$ and a metal with complex permittivity $\epsilon_2$ in the region $z < \zeta(\boldrho)$, where $\boldrho = (x,y,0)$ denotes the in-plane position vector. Restriction is made to non-magnetic media for simplicity. The surface profile, $\zeta(\boldrho)$, is assumed to be an isotropic stationary stochastic process with zero mean $\avg{\zeta (\boldrho)} = 0$ and correlation function $\avg{ \zeta (\boldrho_1) \zeta (\boldrho_2) }  = C(|\boldrho_1 - \boldrho_2|)$, where $\avg{\cdots}$ denotes the average over the ensemble of surface profiles. This assumption aligns well with experimentally measured surface correlation functions \cite{Laverdant2008}.

The electric field for a given realisation of $\zeta$ satisfies the wave equation $\nabla \times \nabla \times \mathbf{E}(\mathbf{r}) - \omega^2 \epsilon(\mathbf{r})\mu_0 \mathbf{E}(\mathbf{r}) = \mathbf{0}$, where $\epsilon(\mathbf{r}) = \epsilon_1 + (\epsilon_2 - \epsilon_1) H(\zeta(\boldrho) - z)$ is the electric permittivity distribution describing the rough surface, $H(z)$ is the Heaviside function and an $e^{-i\omega t}$ time dependence has been assumed. Letting $\epsilon(\mathbf{r}) = {\epsilon}_f(\mathbf{r}) + \delta \epsilon(\mathbf{r})$, the wave equation can be rewritten in the form $\nabla \times \nabla \times \mathbf{E}(\mathbf{r}) - \omega^2 \epsilon_f(\mathbf{r})\mu_0 \mathbf{E}(\mathbf{r}) = \omega^2 \mu_0 \delta \epsilon(\mathbf{r}) \mathbf{E}(\mathbf{r})$. Here $\epsilon_f(\mathbf{r}) =  \epsilon_1 + (\epsilon_2 - \epsilon_1) H(- z) $ describes the permittivity distribution for a flat interface at $z=0$, which upon restricting to the small roughness regime simplifies to $\delta{\epsilon}(\mathbf{r})  \approx (\epsilon_2-\epsilon_1)\delta(z)\zeta(\boldrho) + \mathcal{O}(\zeta^2)$. Formally the electric field distribution $\mathbf{E}(\mathbf{r})$ for a given realisation of surface roughness can then be found using the flat interface Green's tensor $\mathbb{G}(\mathbf{r},\mathbf{r}')$  according to 
\begin{align}
\mathbf{E}(\mathbf{r}) = \mathbf{E}_0(\mathbf{r}) + i\omega \mu_0\int \mathbb{G}(\mathbf{r},\mathbf{r}')\mathbf{J}(\mathbf{r}')d\mathbf{r}'\label{eq:EGreens}
\end{align}
where $\mathbf{E}_0(\mathbf{r})$ is any solution to the homogeneous wave equation for a flat surface and the effect of surface roughness is accounted for by the equivalent current $\mathbf{J}(\mathbf{r}) = -i \omega \delta \epsilon(\mathbf{r}) \mathbf{E}(\mathbf{r})$ \cite{Kroger1970,Bousquet1981}. Physically, $\mathbf{E}_0$ could for example represent an unscattered SPP propagating on a flat surface or an incident (and reflected) optical beam.

Our interest in this work lies in determining the two-point correlations present in roughness induced field fluctuations $\delta \mathbf{E}(\mathbf{r})  = \mathbf{E}(\mathbf{r}) - \avg{\mathbf{E}(\mathbf{r})}$ as described by the  matrix
\begin{align}
\mathbb{W}(\mathbf{r}_1,\mathbf{r}_2) = \avg{ \delta \mathbf{E}^*(\mathbf{r}_1) \delta \mathbf{E}^T(\mathbf{r}_2) }.\label{eq:Wdef}
\end{align}
Within the first order Born approximation (hence neglecting any multiple scattering) we note that the mean field distribution $\avg{\mathbf{E}(\mathbf{r})}$ is simply $ \mathbf{E}_0(\mathbf{r}) $, such that the correlation matrix can be found using Eqs.~\eqref{eq:EGreens} and \eqref{eq:Wdef} and is given by
\begin{align}
\mathbb{W}(\mathbf{r}_1,\mathbf{r}_2)=\omega^2 \mu_0^2  \iint \mathbb{G}^*(\mathbf{r}_1,\mathbf{r}_3) \mathbb{W}_J(\mathbf{r}_3,\mathbf{r}_4)    \mathbb{G}^T(\mathbf{r}_2,\mathbf{r}_4)  d \mathbf{r}_3d \mathbf{r}_4 \label{eq:W}
\end{align}
where $\mathbb{W}_J(\mathbf{r}_1,\mathbf{r}_2) = \avg{\mathbf{{J}^*(\mathbf{r}_1)\mathbf{J}^T(\mathbf{r}_2)}}$ is the two point correlation matrix for the equivalent surface current $\mathbf{J}(\mathbf{r}) = -i \omega \delta \epsilon(\mathbf{r}) \mathbf{E}_0(\mathbf{r})$. Since $\mathbf{E}_0$ is independent of the surface roughness profile by construction, the current correlation matrix can be expressed in the form 
\begin{align}
\mathbb{W}_J(\mathbf{r}_1,\mathbf{r}_2) = \omega^2 |\epsilon_2 - \epsilon_1|^2 \delta(z_1)\delta(z_2) C(P) \mathbf{E}^*_0(\mathbf{r}_1)\mathbf{E}^T_0(\mathbf{r}_2)
\end{align} 
where $\mathbf{P}=\boldsymbol{\rho}_1-\boldsymbol{\rho}_2 = P[\cos\theta,\sin\theta,0]^T$. Hitherto we have only accounted for randomness arising from the surface roughness, however, in reality the reference field $\mathbf{E}_0$ (assumed thus far to be time harmonic), may exhibit stochastic fluctuations in time, space or polarisation due to the nature of the exciting source. Assuming these fluctuations to be statistically independent of $\zeta$ and restricting to planar Schell-model sources \cite{Mandel1995}, we have $\mathbb{W}_J(\mathbf{r}_1,\mathbf{r}_2) = \mathbbold{w}_J(\mathbf{P}) \delta(z_1)\delta(z_2)$ where
\begin{align}
\mathbbold{w}_J(\mathbf{P}) = \omega^2 |\epsilon_2 - \epsilon_1|^2  C(P) \mathbb{W}_0(\mathbf{P}),
\end{align} 
$ \mathbb{W}_0(\mathbf{P}) = \avg{ \mathbf{E}^*_0(\boldrho_1)\mathbf{E}^T_0(\boldrho_2)}$ is the cross-spectral density matrix for the reference system and the ensemble average denoted by $\avg{\cdots}$ now also includes averaging over the ensemble of sources. 

To simplify Eq.~\eqref{eq:W} we note that  the Green's tensor $\mathbb{G}(\mathbf{r}_1,\mathbf{r}_2)$ is only a function of the relative in-plane position vector $\boldrho_1-\boldrho_2$ and  can thus be expressed in the spatial frequency domain by means of a two-dimensional Fourier transform. With a little manipulation (see Supplementary Material) it follows that
\begin{align}
\mathbb{W}(\mathbf{P},z_1,z_2)
& =  \frac{\omega^2\mu_0^2}{(2\pi)^2}\iint d \boldsymbol{\kappa}  \,d\mathbf{Q} \,e^{-i\boldsymbol{\kappa}\cdot(\mathbf{P}-\mathbf{Q})  }  \widetilde{\mathbb{G}}^*(\boldsymbol{\kappa},z_1,0)
\mathbbold{w}_J(\mathbf{Q}) 
\widetilde{\mathbb{G}}^T(\boldsymbol{\kappa},z_2,0) ,\label{eq:Greens_Fourier2}
\end{align}
where we observe that $\mathbb{W}$ is a function of $(\mathbf{P},z_1,z_2)$ only, $\mathbf{Q} = \boldrho_3 - \boldrho_4 = Q[\cos\beta,\sin\beta,0]^T$ and $\boldsymbol{\kappa} = \kappa[\cos\phi,\sin\phi,0]^T$. The Fourier domain Green's function, $\widetilde{\mathbb{G}}(\boldsymbol{\kappa},z_1,z_2)$, for SPP scattering  is given by \cite{Sondergaard2004}
\begin{align}
\widetilde{\mathbb{G}}(\boldsymbol{\kappa},z_1,z_2) &= \frac{4\pi^2\kappa e^{-\alpha\kappa (z_1+z_2)} }{K_0 (\gamma^2 - \kappa^2 )}[\hat{\mathbf{z}} - i \alpha \hat{\boldsymbol{\kappa}}][\hat{\mathbf{z}}  + i \alpha \hat{\boldsymbol{\kappa}}]^T \label{eq:GSPP}
\end{align}
where $K_0 = 2\pi^2 \sqrt{-\epsilon_1\epsilon_2} \left(1-{\epsilon_1^2}/{\epsilon_2^2}\right) {(\epsilon_1 +\epsilon_2)}/ {(\epsilon_1\epsilon_2)}$, $\alpha = \sqrt{-\epsilon_1/\epsilon_2}$, $\gamma = k_0 \sqrt{{\epsilon_1\epsilon_2}/({\epsilon_1 +\epsilon_2})}$ is the SPP propagation constant, $k_0 = \omega / c$, $c$ is the speed of light in vacuum and hat notation is used to denote unit vectors e.g. $\hat{\boldsymbol{\kappa}} = |\boldsymbol{\kappa}| /\kappa$.
Expressing the correlation matrix $\mathbbold{w}_J$ in the dyadic form 
\begin{align}
\mathbbold{w}_J(\mathbf{Q}) = \sum_{i=1}^3\sum_{j=1}^3 w_{ij}(\mathbf{Q}) \hat{\mathbf{e}}_i(\beta) \hat{\mathbf{e}}_j^T(\beta)
\end{align}
where $\hat{\mathbf{e}}_1(\beta) = \hat{\mathbf{Q}} $, $\hat{\mathbf{e}}_2(\beta) = \hat{\boldsymbol{\beta}}$ and $\hat{\mathbf{e}}_3(\beta) =\hat{\mathbf{z}}$, and using Eq.~\eqref{eq:GSPP}, the integration in Eq.~\eqref{eq:Greens_Fourier2} can then be performed using standard integral identities \cite{Stratton1941a}. For isotropic  $\mathbb{W}_0$ we ultimately find
\begin{align}
&\mathbb{W}(\mathbf{P},z_1,z_2)=   \frac{(2\pi)^4\omega^2\mu_0^2  }{|K_0|^2}    \int_0^\infty  Q \,\{|\alpha|^2 w_{11}(Q) +w_{33}(Q)\} \mathbb{V}_{0}(P,Q,\theta,z_1,z_2) dQ\label{eq:WSPPfinal}
\end{align} 
where (dropping the functional dependency for clarity)
\begin{align}
&\mathbb{V}_0 =\frac{1}{2}\left[\begin{array}{ccc} 
|\alpha|^2L_-& -|\alpha|^2L_{02} \sin 2 \theta & 2\alpha^* L_{01} \cos \theta \\
-|\alpha|^2 L_{02} \sin 2 \theta  &|\alpha|^2L_+& 2\alpha^* L_{01} \sin \theta \\
-2\alpha L_{01} \cos \theta & - 2\alpha L_{01}  \sin \theta & 2L_{00} 
\end{array}\right]
\end{align}
and $L_{\pm} = L_{00} \pm L_{02} \cos 2 \theta$. The $L_{nm}(P,Q,z_1,z_2)$ terms denote integrals over ${\kappa}$ as detailed in the Supplementary Material. Closed form expressions for $L_{nm}$ can however be found using Jordan's lemma and  Cauchy's residue theorem \cite{Morse1953}, yielding $L_{nm}= l_{nm}(\gamma)  +l_{nm}(-\gamma^*)$, where
\begin{align}
l_{nm}(\kappa) \!=\! \frac{\pi}{4}\frac{\sigma^{n+m}\kappa^2 e^{-\sigma\kappa \mathcal{Z}}}{\mbox{Im}[\kappa^2] }\!\left\{  \begin{array}{ll} 
\!\!J_n(\kappa Q)H_m^{(1)}(\kappa P) &\!\!\text{for~} P > Q \\
\!\!H_n^{(1)}(\kappa Q ) J_m(\kappa P) &\!\! \text{for~} P < Q
\end{array}\right. \label{eq:Lmn}
\end{align}
and $\mathcal{Z} = \alpha^* z_1 + \alpha z_2$ and $\sigma(\kappa) = \text{sign}[\mbox{Re}(\kappa)]$. 

Eqs.~\eqref{eq:WSPPfinal}--\eqref{eq:Lmn} represent the central analytic result of this article.  It is important to note, however, that the form of admissible $\mathbb{W}_0$ is limited through the assumption of in-plane isotropy, i.e. rotational invariance. Specifically $\mathbb{W}_0(\mathbf{P})$ must have the form
\begin{align}
\mathbb{W}_0(\mathbf{P}) = \left[\begin{array}{ccc}
w_{11}(\mathbf{P}) & w_{12}(\mathbf{P}) & 0 \\
-w_{12}(\mathbf{P}) & w_{11}(\mathbf{P}) & 0 \\
0 & 0 & w_{33}(\mathbf{P})
\end{array}\right]
\end{align}
of which normally incident circularly polarised and unpolarised light are important examples. Although an angular dependence can be introduced into the form of $\mathbbold{w}_J$ so as to allow a wider class of background fields to be described (ultimately giving rise to the presence of higher order Bessel functions in Eq.~\eqref{eq:WSPPfinal}), we have here elected to consider the most symmetric case so as not to overly complicate the mathematics. As evident from Eq.~\eqref{eq:WSPPfinal}, the precise nature of correlations in the roughness induced field fluctuations is  seen to be dependent on the source and surface correlation functions through $w_{11}$ and $w_{33}$, however they are insensitive to the off-diagonal elements $w_{12}$. 

\section*{Results and discussion}

With Eqs.~\eqref{eq:WSPPfinal}-\eqref{eq:Lmn} in hand, the remainder of this article focuses on studying some general and limiting properties of $\mathbb{W}$ in more detail. Our analysis begins with consideration of the polarisation properties of the plasmon speckle pattern. Since we consider an SPP field which both evanescently decays with distance from the surface and contains non-planar wavefronts due to random scattering, it is necessary to adopt a full three-dimensional (3D) formalism \cite{Setala2002,Tervo2003} when discussing polarisation properties. In particular we consider the 3D spectral degree of cross polarisation (DOCP), $\mathcal{D}_{\txtpow{3D}}(\mathbf{P})$, defined by \cite{Volkov2008,Setala2002}
\begin{align}
\!\mathcal{D}_{n\txtpow{D}}^2(\mathbf{P},z_1,z_2) = \left(3-\frac{n}{2}\right)\left[ \frac{\|\mathbb{W}(\mathbf{P},z_1,z_2)\|_F^2}{\text{tr}[\mathbb{W}(\mathbf{P},z_1,z_2)]^2} -\frac{1}{n} \right]\label{eq:DOPdef}
\end{align}
where $\|\cdots\|_F$ denotes the Frobenius matrix norm. Eq.~\eqref{eq:DOPdef} also defines the two-dimensional (2D) DOCP ($n = 2$) as will be discussed below. 

Initially restricting attention to $\mathbf{P}=\mathbf{0}$, whereby the DOCP is equivalent to the spectral degree of polarisation (DOP), we note that  $L_{nm}(P=0,Q,z_1,z_2) \sim \delta_{m0}$. It therefore follows that $\mathbb{V}_0(0,Q,\theta,z_1,z_2) =  L_{00} \mathbb{A}$ and $\mathbb{W}(\mathbf{0},z_1,z_2) \sim \mathbb{A}$ where $\mathbb{A} = \text{diag}[|\alpha|^2,|\alpha|^2,2]$ is a constant diagonal matrix. Accordingly the DOCP  is
\begin{align}
\mathcal{D}_{\txtpow{3D}}(\mathbf{0},z_1,z_2) = \frac{|\alpha|^2 -2}{2(1+|\alpha|^2 )}\label{eq:DOP3}
\end{align}
which is independent of the axial position.
We can similarly consider the 2D DOCP at $\mathbf{P}=\mathbf{0}$ defined with respect to correlations between different in-plane field components ($E_x$ and $E_y$). By virtue of the assumed rotational invariance within the $x$-$y$ plane, this is equivalent to correlations between the radial and azimuthal field components $E_\rho$ and $E_\theta$). Moreover we can consider correlations between the axial and in-plane components. The associated $2\times 2$ correlation matrices $\mathbb{W}_{\parallel}$ and $\mathbb{W}_{\perp}$ can be found by extracting the correct block matrices from $\mathbb{W}$. We find that $\mathcal{D}_{\txtpow{2D}}^{\parallel} = 0$, showing that the in-plane field is unpolarised as would be expected from the assumed symmetry. Since SPPs on a flat surface are TM modes there is however always a non-zero out of plane field component in turn yielding a non-zero 2D DOCP, $\mathcal{D}_{\txtpow{2D}}^{\perp}$, satisfying $3[\mathcal{D}_{\txtpow{2D}}^{\perp}(\mathbf{0})]^2 = 4[\mathcal{D}_{\txtpow{3D}}(\mathbf{0})]^2 - 1 $.

\begin{figure}[t!]
	\begin{center}
		\includegraphics[width=0.5\columnwidth]{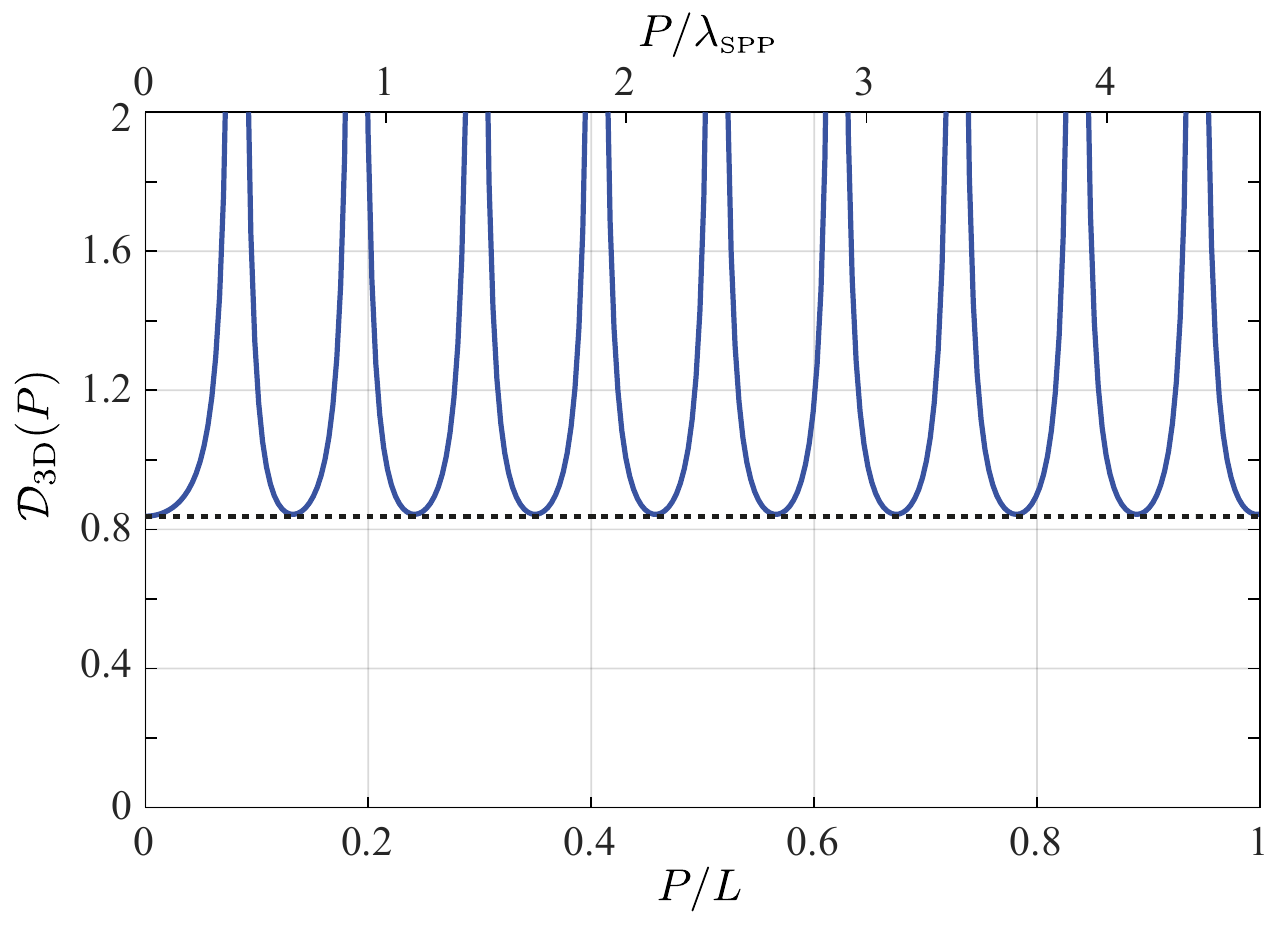}
		\caption{Degree of cross-polarisation, $\mathcal{D}_{3\txtpow{D}}$, for a SPP speckle field supported on a rough surface with Gaussian surface correlation function of correlation length $Q_0 = \lambda_{\txtpow{SPP}}/4$. Separation coordinate has been normalised to both the SPP attenuation length $L = 1/2\mbox{Im}[\gamma]$ (bottom axis) and SPP wavelength (top axis). Horizontal dashed line corresponds to limiting value given by Eq.~\eqref{eq:DOP3}.\label{fig:DOP}}
	\end{center}
\end{figure}

An example of the behaviour of the spectral DOCP for two separated co-planar points ($P>0$, $z_1 = z_2 = 0$) is shown in  Figure~\ref{fig:DOP} assuming a rough gold-air interface ($\epsilon_1 = 1$, $\epsilon_2 = -7.99 + 2.06i$ at a free-space wavelength of 600~nm \cite{Rakic1998}), normally incident circularly polarised light and a Gaussian surface correlation function $C(Q) = h^2 \exp[-Q^2 	/(2Q_0^2)]$ of width $Q_0 = 0.25\lambda_{\txtpow{SPP}}$, where $h^2$ is the  mean square height deviation of the surface and $\lambda_{\txtpow{SPP}} = 2\pi/\mbox{Re}[\gamma]$ is the SPP wavelength. We note that the DOCP is independent of $h$ since it appears as a common factor in the numerator and denominator in Eq.~\eqref{eq:DOPdef} and hence cancels. It is evident from Figure~\ref{fig:DOP} that the zero-separation DOCP (given by Eq.~\eqref{eq:DOP3} and shown by the dashed line) acts as a lower bound on $\mathcal{D}_{3\txtpow{D}}$ for all separations $P\geq 0$.
Moreover the DOCP regularly takes this limiting value with a period given by the SPP propagation wavelength. We also note that since $\mathbb{W}(\mathbf{P},z_1,z_2)$ is not non-negative definite for $\mathbf{P}\neq \mathbf{0}$, the 3D DOCP can adopt values greater than unity in contrast to the spectral DOP $\mathcal{D}_{3\txtpow{D}}(\mathbf{0})$ \cite{Volkov2008}. Infinite values are indeed seen in Figure~\ref{fig:DOP} corresponding to separations for which the points have zero degree of coherence (see below), i.e., where the denominator in Eq.~\eqref{eq:DOPdef} is zero. 

A further common metric parametrising random fields is the so-called spectral degree of coherence (DOC), which provides a measure of fringe visibility in interferometric measurements and is thus pertinent to the growing number of plasmon interferometers \cite{Gao2013,Feng2012a}. The DOC is defined here with respect to the correlation matrix at $\boldrho_1 = \boldrho_2 = \mathbf{0}$ and we restrict to the $z_1 = z_2 = z_0$ plane such that $\mu(\mathbf{P},z_0) =\tr[\mathbb{W}(\mathbf{P},z_0,z_0)] /\tr[\mathbb{W}(\mathbf{0},0,0)]$ \cite{Setala2003a}. To evaluate the DOC we first note that $\text{tr}[\mathbb{V}_0] = (|\alpha|^2 + 1)L_{00}$ and partition the integration domain in Eq.~\eqref{eq:GSPP} into the two regions defined by $0 \leq Q < P$ and $ P \leq Q \leq \infty$. 
The general form of the DOC, given in the Supplementary Material, is dictated by the functional form of the surface and source correlation functions through $w_{11}$ and $w_{33}$. In contrast to the universal, i.e. surface and source independent, characteristics of speckle patterns in the far field \cite{Ponomarenko2002}, non-universal correlations are thus exhibited in random SPP scattering. Similar non-universal behaviour in near field speckle patterns has been previously discussed \cite{Greffet1995a}, and may be useful for imaging or sensing \cite{Skipetrov2000,Dogariu2015}.  Noting, however, that typically $\mbox{Im}[\gamma] \ll \mbox{Re}[\gamma]$, that is to say the SPP propagation length ($L=1/2\mbox{Im}[\gamma]$) is much larger than the wavelength, we can approximate the DOC according to 
\begin{align}
&\!\!\!\mu(\mathbf{P}) \approx\frac{e^{-\gamma \mathcal{Z}_0 }}{2}  \Big\{\Big[ H^{(1)}_0(\gamma P) - H^{(1)}_0(-\gamma^* P) \Big] f_J^P  + J_0(\gamma P)[f_J^\infty + i f_Y^\infty] + J_0(-\gamma^* P)[f_J^\infty- if_Y^\infty] \Big\} , \label{eq:mulowloss}
\end{align}
where the $f$ factors describe the ratio of integrals defined in the Supplementary Material and $\mathcal{Z}_0 = 2\mbox{Re}[\alpha] z_0$. For our purposes here, the precise form of the integrals is unimportant except to note that the integrand is dependent on $w_{11}$ and $w_{33}$. 

\begin{figure}[t!]
	\begin{center}
		\includegraphics[width=0.5\columnwidth]{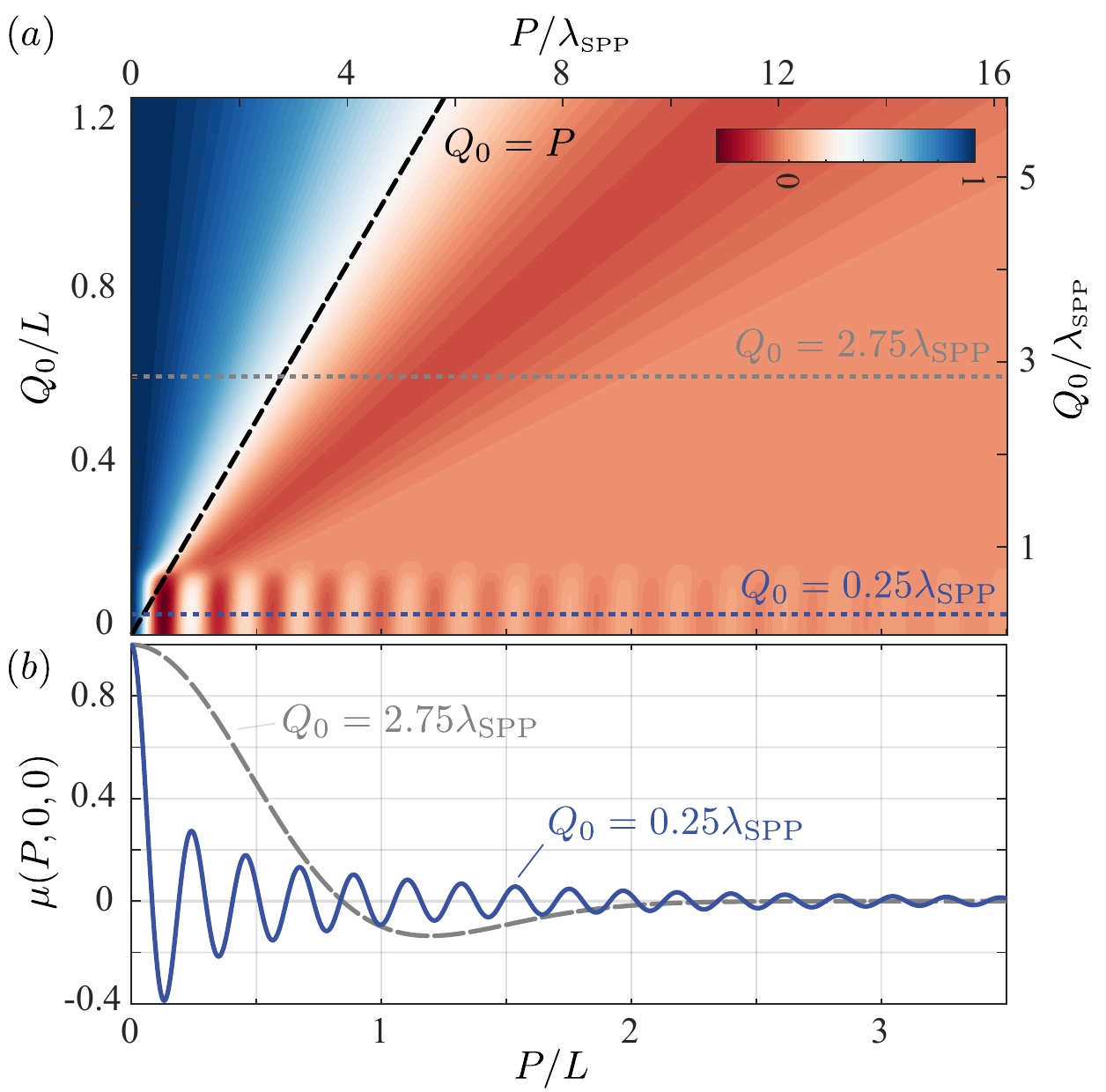}
		\caption{(a) Degree of coherence $\mu(P,0,0)$ as a function of surface correlation length $Q_0$. Horizontal dashed lines correspond to cross-sections plotted in (b).}\label{fig:DOC}
	\end{center}
\end{figure}

A number of conclusions can be made on the basis of Eq.~\eqref{eq:mulowloss}. Firstly, for a loss-free metal $\epsilon_2$ is purely real and negative and consequently the SPP propagation constant $\gamma$ is real. Using the reflection formulae for the Bessel and Hankel functions \cite{Abramowitz1972} it follows that the DOC  reduces to the universal form $\mu(\mathbf{P}) = J_0(\gamma P)$ which decays as $\sim P^{-1/2}$. The Bessel function dependence of the DOC is familiar from the case of 2D scalar waves, however, we note we have derived it here using the full electromagnetic expressions for SPP waves. Physically, however, all metals exhibit some loss and it is not reasonable to neglect absorption. Instead we observe that typical surface and/or source correlation functions decay over some characteristic correlation length $Q_0$. When considering the DOC between points separated by distances greater than $Q_0$ length, by virtue of the dependence of the $f$ terms on ${w}_{11}$ and $w_{33}$ it can be shown that $f_J^\infty \approx f_Y^\infty \approx 0$ and $f_J^P  \approx 1$, such that 
\begin{align}
\mu(\mathbf{P},z_0) \approx \frac{e^{-\gamma \mathcal{Z}_0 }}{2}\left[H^{(1)}_0(\gamma P) - H^{(1)}_0(-\gamma^* P) \right] .\label{eq:DOCHankel}
\end{align}	
This analytic form is notably independent of the precise nature of the correlations $\mathbbold{w}_J$.
For separations $P\lesssim Q_0$, however, non-universal behaviour of $\mu$ is manifest.
Nevertheless in the limit that $Q_0 \rightarrow 0$, corresponding to an infinitesimally short surface correlation length, or use of a thermal source \cite{Henkel2000}, the DOC is described by Eq.~\eqref{eq:DOCHankel} for all $P>0$.  This behaviour can be seen in Figure~\ref{fig:DOC}(a), where we have plotted the DOC calculated numerically from Eq.~\eqref{eq:WSPPfinal} as a function of both the spatial separation $P$ and the surface correlation length $Q_0$. All other simulation parameters are the same as in Figure~\ref{fig:DOP}. Specifically we note that for correlation lengths $Q_0 \lesssim \lambda_{\txtpow{SPP}} /2$ the functional behaviour described by Eq.~\eqref{eq:DOCHankel} is apparent, whereas for correlation lengths larger than $\sim\lambda_{\txtpow{SPP}}$ the DOC is dictated by the correlation function $\mathbbold{w}_J$. Whilst in this latter case Eq.~\eqref{eq:DOCHankel} holds at large separations, the DOC has nevertheless fallen essentially to zero such that the SPP oscillations are not visible. Cross-sections of $\mu(P,0,0)$ for $Q_0 = 0.25\lambda_{\txtpow{SPP}}$ and $2.75\lambda_{\txtpow{SPP}}$ (as depicted by the horizontal dashed lines in Figure~\ref{fig:DOC}(a)) are also shown in Figure~\ref{fig:DOC}(b), which demonstrate this differing behaviour. 
Finally, we observe that there exists a lower limit to the scale of fluctuations in SPP fields as can be seen in Figure~\ref{fig:DOC} and is captured in Eq.~\eqref{eq:DOCHankel}. In particular, the minimum fluctuation length is on the order of the surface plasmon wavelength. This behaviour is notably different to that of near field speckle patterns comprised of evanescent waves for which sub-wavelength correlation lengths can occur \cite{Carminati2010}. The difference in behaviour can be simply understood by noting that in near field speckle, a range of waves with differing transverse wavenumbers ${\kappa}$ contribute to the total field, whereas in SPP speckle the resonant nature of SPPs means only a single transverse wavenumber is present and thus the the frequency of fluctuations in lower bounded.

\section*{Conclusions}
In summary, this work has considered field correlations present in surface plasmon speckle patterns arising from scattering of SPPs propagating on a rough surface. Within the first order Born approximation and assuming in-plane statistical isotropy and homogeneity, we have derived closed form analytic expressions for the correlation matrix for roughness induced field fluctuations relative to a smooth metallo-dielectric interface, as given by Eqs.~\eqref{eq:WSPPfinal}--\eqref{eq:Lmn}. Correlations were thereby found to be independent of off-diagonal elements of the source spectral density matrix and to exponentially decay with distance from the interface. Analytic formulae derived also allowed us to investigate the polarisation and coherence properties of SPP speckle patterns in closer detail. In particular, we found that the 3D DOP is not only independent of the axial distance from the interface, but that it also serves as a lower bound for the DOCP across the whole speckle distribution. Universal forms for the DOC were also shown to arise for low loss metals or when the correlation length of the surface height, or source field were small ($< \lambda_{\txtpow{SPP}}/2$). Non-universal behaviour can however result when such conditions are not met whereby the DOC (and also the  correlation matrix $\mathbb{W}$) is dominated by the source/surface correlation function. A minimum fluctuation length, on the order of the SPP wavelength, was thus found in contrast to purely evanescent near field speckle patterns. 

\section*{Supplementary Material}
See supplementary material for full derivations.

\section*{Acknowledgements}
This work was funding by the Royal Society.


\section*{Acknowledgements}

This work was funded by the Royal Society through a University Research Fellowship. 

\section*{Author contributions statement}

MRF conceived the idea, performed derivations, numerical calculations and wrote the article.

\section*{Additional information}

\textbf{Competing interests:} The author declares no competing interests.


\begin{thebibliography}{10}
	\urlstyle{rm}
	\expandafter\ifx\csname url\endcsname\relax
	\def\url#1{\texttt{#1}}\fi
	\expandafter\ifx\csname urlprefix\endcsname\relax\def\urlprefix{URL }\fi
	\expandafter\ifx\csname doiprefix\endcsname\relax\def\doiprefix{DOI: }\fi
	\providecommand{\bibinfo}[2]{#2}
	\providecommand{\eprint}[2][]{\url{#2}}
	
	\bibitem{Sheng1995}
	\bibinfo{author}{Sheng, P.}
	\newblock \emph{\bibinfo{title}{{Introduction to Wave Scattering, Localization
				and Mesoscopic Phenomena}}} (\bibinfo{publisher}{Academic Press Inc., New
		York}, \bibinfo{year}{1995}).
	
	\bibitem{vanRossum1998}
	\bibinfo{author}{van Rossum, M. C.~W.} \& \bibinfo{author}{Nieuwenhuizen,
		T.~M.}
	\newblock \bibinfo{journal}{\bibinfo{title}{{Multiple scattering of classical
				waves: from microscopy to mesoscopy and diffusion}}}.
	\newblock {\emph{\JournalTitle{Rev. Mod. Phys.}}}
	\textbf{\bibinfo{volume}{71}}, \bibinfo{pages}{313--371},
	\doiprefix\url{10.1103/RevModPhys.71.313} (\bibinfo{year}{1999}).
	
	\bibitem{Berkovitsa1994}
	\bibinfo{author}{Berkovits, R.} \& \bibinfo{author}{Feng, S.}
	\newblock \bibinfo{journal}{\bibinfo{title}{{Correlations in coherent multiple
				scattering}}}.
	\newblock {\emph{\JournalTitle{Phys. Reports}}} \textbf{\bibinfo{volume}{238}},
	\bibinfo{pages}{135--172}, \doiprefix\url{10.1016/0370-1573(94)90079-5}
	(\bibinfo{year}{1994}).
	
	\bibitem{Goodman2006}
	\bibinfo{author}{Goodman, J.~W.}
	\newblock \emph{\bibinfo{title}{{Speckle Phenomena in Optics: Theory and
				Applications}}} (\bibinfo{publisher}{Roberts and Company Publishers},
	\bibinfo{address}{Englewood, Colorado}, \bibinfo{year}{2006}).
	
	\bibitem{Dogariu2015}
	\bibinfo{author}{Dogariu, A.} \& \bibinfo{author}{Carminati, R.}
	\newblock \bibinfo{journal}{\bibinfo{title}{{Electromagnetic field correlations
				in three-dimensional speckles}}}.
	\newblock {\emph{\JournalTitle{Phys. Rep.}}} \textbf{\bibinfo{volume}{559}},
	\bibinfo{pages}{1--29}, \doiprefix\url{10.1016/j.physrep.2014.11.003}
	(\bibinfo{year}{2015}).
	
	\bibitem{Feng1988}
	\bibinfo{author}{Feng, S.}, \bibinfo{author}{Kane, C.}, \bibinfo{author}{Lee,
		P.~A.} \& \bibinfo{author}{Stone, A.~D.}
	\newblock \bibinfo{journal}{\bibinfo{title}{{Correlations and fluctuations of
				coherent wave transmission through disordered media}}}.
	\newblock {\emph{\JournalTitle{Phys. Rev. Lett.}}}
	\textbf{\bibinfo{volume}{61}}, \bibinfo{pages}{834--837},
	\doiprefix\url{10.1103/PhysRevLett.61.834} (\bibinfo{year}{1988}).
	
	\bibitem{Shnerb1991}
	\bibinfo{author}{Shnerb, N.} \& \bibinfo{author}{Kaveh, M.}
	\newblock \bibinfo{journal}{\bibinfo{title}{{Non-Rayleigh statistics of waves
				in random systems}}}.
	\newblock {\emph{\JournalTitle{Phys. Rev. B}}} \textbf{\bibinfo{volume}{43}},
	\bibinfo{pages}{1279--1282}, \doiprefix\url{10.1103/PhysRevB.43.1279}
	(\bibinfo{year}{1991}).
	
	\bibitem{Kogan1993}
	\bibinfo{author}{Kogan, E.}, \bibinfo{author}{Kaveh, M.},
	\bibinfo{author}{Baumgartner, R.} \& \bibinfo{author}{Berkovits, R.}
	\newblock \bibinfo{journal}{\bibinfo{title}{{Statistics of waves propagating in
				a random medium}}}.
	\newblock {\emph{\JournalTitle{Phys. Rev. B}}} \textbf{\bibinfo{volume}{48}},
	\bibinfo{pages}{9404--9410}, \doiprefix\url{10.1103/PhysRevB.48.9404}
	(\bibinfo{year}{1993}).
	
	\bibitem{Skipetrov2000}
	\bibinfo{author}{Skipetrov, S.~E.} \& \bibinfo{author}{Maynard, R.}
	\newblock \bibinfo{journal}{\bibinfo{title}{{Nonuniversal correlations in
				multiple scattering}}}.
	\newblock {\emph{\JournalTitle{Phys. Rev. B}}} \textbf{\bibinfo{volume}{62}},
	\bibinfo{pages}{886--891}, \doiprefix\url{10.1103/PhysRevB.62.886}
	(\bibinfo{year}{2000}).
	
	\bibitem{Bertolotti2012}
	\bibinfo{author}{Bertolotti, J.} \emph{et~al.}
	\newblock \bibinfo{journal}{\bibinfo{title}{{Non-invasive imaging through
				opaque scattering layers}}}.
	\newblock {\emph{\JournalTitle{Nature}}} \textbf{\bibinfo{volume}{491}},
	\bibinfo{pages}{232--234}, \doiprefix\url{10.1038/nature11578}
	(\bibinfo{year}{2012}).
	
	\bibitem{Berne2000}
	\bibinfo{author}{Berne, B.~J.} \& \bibinfo{author}{Pecora, R.}
	\newblock \emph{\bibinfo{title}{{Dynamic Light Scattering With Applications to
				Chemistry, Biology, and Physics}}} (\bibinfo{publisher}{Dover Publications,
		New York}, \bibinfo{year}{2000}).
	
	\bibitem{Lemoult2009}
	\bibinfo{author}{Lemoult, F.}, \bibinfo{author}{Lerosey, G.},
	\bibinfo{author}{{De Rosny}, J.} \& \bibinfo{author}{Fink, M.}
	\newblock \bibinfo{journal}{\bibinfo{title}{{Manipulating Spatiotemporal
				Degrees of Freedom of Waves in Random Media}}}.
	\newblock {\emph{\JournalTitle{Phys. Rev. Lett.}}}
	\textbf{\bibinfo{volume}{103}}, \bibinfo{pages}{173902},
	\doiprefix\url{10.1103/PhysRevLett.103.173902} (\bibinfo{year}{2009}).
	
	\bibitem{Bender2017}
	\bibinfo{author}{Bender, N.}, \bibinfo{author}{Yilmaz, H.},
	\bibinfo{author}{Bromberg, Y.} \& \bibinfo{author}{Cao, H.}
	\newblock \bibinfo{journal}{\bibinfo{title}{{Customizing Speckle Intensity
				Statistics}}}.
	\newblock {\emph{\JournalTitle{Optica}}} \textbf{\bibinfo{volume}{5}},
	\bibinfo{pages}{595--600}, \doiprefix\url{10.1364/FIO.2017.FW5B.2}
	(\bibinfo{year}{2017}).
	
	\bibitem{Apostol2003}
	\bibinfo{author}{Apostol, A.} \& \bibinfo{author}{Dogariu, A.}
	\newblock \bibinfo{journal}{\bibinfo{title}{{Spatial correlations in the near
				field of random media}}}.
	\newblock {\emph{\JournalTitle{Phys. Rev. Lett.}}}
	\textbf{\bibinfo{volume}{91}}, \bibinfo{pages}{93901},
	\doiprefix\url{10.1103/PhysRevLett.91.093901} (\bibinfo{year}{2003}).
	
	\bibitem{Carminati2010}
	\bibinfo{author}{Carminati, R.}
	\newblock \bibinfo{journal}{\bibinfo{title}{{Subwavelength spatial correlations
				in near-field speckle patterns}}}.
	\newblock {\emph{\JournalTitle{Phys. Rev. A}}} \textbf{\bibinfo{volume}{81}},
	\bibinfo{pages}{053804}, \doiprefix\url{10.1103/PhysRevA.81.053804}
	(\bibinfo{year}{2010}).
	
	\bibitem{Liu2005f}
	\bibinfo{author}{Liu, C.} \& \bibinfo{author}{Park, S.-H.}
	\newblock \bibinfo{journal}{\bibinfo{title}{{Anisotropy of near-field speckle
				patterns.}}}
	\newblock {\emph{\JournalTitle{Opt. Lett.}}} \textbf{\bibinfo{volume}{30}},
	\bibinfo{pages}{1602--1604}, \doiprefix\url{10.1364/OL.30.001602}
	(\bibinfo{year}{2005}).
	
	\bibitem{Laverdant2008}
	\bibinfo{author}{Laverdant, J.}, \bibinfo{author}{Buil, S.},
	\bibinfo{author}{B{\'{e}}rini, B.} \& \bibinfo{author}{Qu{\'{e}}lin, X.}
	\newblock \bibinfo{journal}{\bibinfo{title}{{Polarization dependent near-field
				speckle of random gold films}}}.
	\newblock {\emph{\JournalTitle{Phys. Rev. B}}} \textbf{\bibinfo{volume}{77}},
	\bibinfo{pages}{165406}, \doiprefix\url{10.1103/PhysRevB.77.165406}
	(\bibinfo{year}{2008}).
	
	\bibitem{Parigi2016a}
	\bibinfo{author}{Parigi, V.} \emph{et~al.}
	\newblock \bibinfo{journal}{\bibinfo{title}{{Near-field to far-field
				characterization of speckle patterns generated by disordered
				nanomaterials}}}.
	\newblock {\emph{\JournalTitle{Opt. Express}}} \textbf{\bibinfo{volume}{24}},
	\bibinfo{pages}{7019--7027}, \doiprefix\url{10.1364/OE.24.007019}
	(\bibinfo{year}{2016}).
	
	\bibitem{Greffet1995a}
	\bibinfo{author}{Greffet, J.-J.} \& \bibinfo{author}{Carminati, R.}
	\newblock \bibinfo{journal}{\bibinfo{title}{{Relationship between the
				near-field speckle pattern and the statistical properties of a surface}}}.
	\newblock {\emph{\JournalTitle{Ultramicros.}}} \textbf{\bibinfo{volume}{61}},
	\bibinfo{pages}{43--50}, \doiprefix\url{10.1016/0304-3991(95)00101-8}
	(\bibinfo{year}{1995}).
	
	\bibitem{Caze2010}
	\bibinfo{author}{Caz{\'{e}}, A.}, \bibinfo{author}{Pierrat, R.} \&
	\bibinfo{author}{Carminati, R.}
	\newblock \bibinfo{journal}{\bibinfo{title}{{Near-field interactions and
				nonuniversality in speckle patterns produced by a point source in a
				disordered medium}}}.
	\newblock {\emph{\JournalTitle{Phys. Rev. A}}} \textbf{\bibinfo{volume}{82}},
	\bibinfo{pages}{043823}, \doiprefix\url{10.1103/PhysRevA.82.043823}
	(\bibinfo{year}{2010}).
	
	\bibitem{Nussenzveig1987}
	\bibinfo{author}{Nussenzveig, H.~M.}, \bibinfo{author}{Foley, J.~T.},
	\bibinfo{author}{Kim, K.} \& \bibinfo{author}{Wolf, E.}
	\newblock \bibinfo{journal}{\bibinfo{title}{{Field correlations within a
				fluctuating homogeneous medium}}}.
	\newblock {\emph{\JournalTitle{Phys. Rev. Lett.}}}
	\textbf{\bibinfo{volume}{58}}, \bibinfo{pages}{218--221},
	\doiprefix\url{10.1103/PhysRevLett.58.218} (\bibinfo{year}{1987}).
	
	\bibitem{Ponomarenko2002}
	\bibinfo{author}{Ponomarenko, S.~A.} \& \bibinfo{author}{Wolf, E.}
	\newblock \bibinfo{journal}{\bibinfo{title}{{Universal structure of field
				correlations within a fluctuating medium}}}.
	\newblock {\emph{\JournalTitle{Phys. Rev. E}}} \textbf{\bibinfo{volume}{65}},
	\bibinfo{pages}{016602}, \doiprefix\url{10.1103/PhysRevE.65.016602}
	(\bibinfo{year}{2002}).
	
	\bibitem{Takatori2017}
	\bibinfo{author}{Takatori, K.}, \bibinfo{author}{Okamoto, T.},
	\bibinfo{author}{Ishibashi, K.} \& \bibinfo{author}{Micheletto, R.}
	\newblock \bibinfo{journal}{\bibinfo{title}{{Surface exciton polaritons
				supported by a J-aggregate-dye/air interface at room temperature}}}.
	\newblock {\emph{\JournalTitle{Opt. Lett.}}} \textbf{\bibinfo{volume}{42}},
	\bibinfo{pages}{3876--3879}, \doiprefix\url{10.1364/OL.42.003876}
	(\bibinfo{year}{2017}).
	
	\bibitem{LeGall1997}
	\bibinfo{author}{{Le Gall}, J.}, \bibinfo{author}{Olivier, M.} \&
	\bibinfo{author}{Greffet, J.-J.}
	\newblock \bibinfo{journal}{\bibinfo{title}{{Experimental and theoretical study
				of reflection and coherent thermal emission by a SiC grating supporting a
				surface-phonon polariton}}}.
	\newblock {\emph{\JournalTitle{Phys. Rev. B}}} \textbf{\bibinfo{volume}{55}},
	\bibinfo{pages}{10105--10114}, \doiprefix\url{10.1103/PhysRevB.55.10105}
	(\bibinfo{year}{1997}).
	
	\bibitem{Barnes2003}
	\bibinfo{author}{Barnes, W.~L.}, \bibinfo{author}{Dereux, A.} \&
	\bibinfo{author}{Ebbesen, T.~W.}
	\newblock \bibinfo{journal}{\bibinfo{title}{{Surface plasmon subwavelength
				optics}}}.
	\newblock {\emph{\JournalTitle{Nature}}} \textbf{\bibinfo{volume}{424}},
	\bibinfo{pages}{824--830}, \doiprefix\url{10.1038/nature01937}
	(\bibinfo{year}{2003}).
	
	\bibitem{Shchegrov2000}
	\bibinfo{author}{Shchegrov, A.~V.}, \bibinfo{author}{Joulain, K.},
	\bibinfo{author}{Carminati, R.} \& \bibinfo{author}{Greffet, J.-J.}
	\newblock \bibinfo{journal}{\bibinfo{title}{{Near-Field Spectral Effects due to
				Electromagnetic Surface Excitations}}}.
	\newblock {\emph{\JournalTitle{Phys. Rev. Lett.}}}
	\textbf{\bibinfo{volume}{85}}, \bibinfo{pages}{1548--1551},
	\doiprefix\url{10.1103/PhysRevLett.85.1548} (\bibinfo{year}{2000}).
	
	\bibitem{Gadenne1989}
	\bibinfo{author}{Gadenne, P.}, \bibinfo{author}{Yagil, Y.} \&
	\bibinfo{author}{Deutscher, G.}
	\newblock \bibinfo{journal}{\bibinfo{title}{{Transmittance and reflectance in
				situ measurements of semicontinuous gold films during deposition}}}.
	\newblock {\emph{\JournalTitle{J. Appl. Phys.}}} \textbf{\bibinfo{volume}{66}},
	\bibinfo{pages}{3019--3025}, \doiprefix\url{10.1063/1.344187}
	(\bibinfo{year}{1989}).
	
	\bibitem{Seal2005}
	\bibinfo{author}{Seal, K.} \emph{et~al.}
	\newblock \bibinfo{journal}{\bibinfo{title}{{Near-Field Intensity Correlations
				in Semicontinuous Metal-Dielectric Films}}}.
	\newblock {\emph{\JournalTitle{Phys. Rev. Lett.}}}
	\textbf{\bibinfo{volume}{94}}, \bibinfo{pages}{226101},
	\doiprefix\url{10.1103/PhysRevLett.94.226101} (\bibinfo{year}{2005}).
	
	\bibitem{VanBeijnum2012}
	\bibinfo{author}{van Beijnum, F.}, \bibinfo{author}{Sirre, J.},
	\bibinfo{author}{R{\'{e}}tif, C.} \& \bibinfo{author}{van Exter, M.~P.}
	\newblock \bibinfo{journal}{\bibinfo{title}{{Speckle correlation functions
				applied to surface plasmons}}}.
	\newblock {\emph{\JournalTitle{Phys. Rev. B}}} \textbf{\bibinfo{volume}{85}},
	\bibinfo{pages}{035437}, \doiprefix\url{10.1103/PhysRevB.85.035437}
	(\bibinfo{year}{2012}).
	
	\bibitem{Mandel1995}
	\bibinfo{author}{Mandel, L.} \& \bibinfo{author}{Wolf, E.}
	\newblock \emph{\bibinfo{title}{{Optical Coherence and Quantum Optics}}}
	(\bibinfo{publisher}{Cambridge University Press, Cambridge},
	\bibinfo{year}{1995}).
	
	\bibitem{Setala2002a}
	\bibinfo{author}{Set{\"{a}}l{\"{a}}, T.}, \bibinfo{author}{Kaivola, M.} \&
	\bibinfo{author}{Friberg, A.~T.}
	\newblock \bibinfo{journal}{\bibinfo{title}{{Degree of Polarization in Near
				Fields of Thermal Sources: Effects of Surface Waves}}}.
	\newblock {\emph{\JournalTitle{Phys. Rev. Lett.}}}
	\textbf{\bibinfo{volume}{88}}, \bibinfo{pages}{123902},
	\doiprefix\url{10.1103/PhysRevLett.88.123902} (\bibinfo{year}{2002}).
	
	\bibitem{Carminati1999}
	\bibinfo{author}{Carminati, R.} \& \bibinfo{author}{Greffet, J.-J.}
	\newblock \bibinfo{journal}{\bibinfo{title}{{Near-Field Effects in Spatial
				Coherence of Thermal Sources}}}.
	\newblock {\emph{\JournalTitle{Phys. Rev. Lett.}}}
	\textbf{\bibinfo{volume}{82}}, \bibinfo{pages}{1660--1663},
	\doiprefix\url{10.1103/PhysRevLett.82.1660} (\bibinfo{year}{1999}).
	
	\bibitem{Henkel2000}
	\bibinfo{author}{Henkel, C.}, \bibinfo{author}{Joulain, K.},
	\bibinfo{author}{Carminati, R.} \& \bibinfo{author}{Greffet, J.-J.}
	\newblock \bibinfo{journal}{\bibinfo{title}{{Spatial coherence of thermal near
				fields}}}.
	\newblock {\emph{\JournalTitle{Opt. Commun.}}} \textbf{\bibinfo{volume}{186}},
	\bibinfo{pages}{57--67}, \doiprefix\url{10.1016/S0030-4018(00)01048-8}
	(\bibinfo{year}{2000}).
	
	\bibitem{Caze2013a}
	\bibinfo{author}{Caz{\'{e}}, A.}, \bibinfo{author}{Pierrat, R.} \&
	\bibinfo{author}{Carminati, R.}
	\newblock \bibinfo{journal}{\bibinfo{title}{{Spatial Coherence in Complex
				Photonic and Plasmonic Systems}}}.
	\newblock {\emph{\JournalTitle{Phys. Rev. Lett.}}}
	\textbf{\bibinfo{volume}{110}}, \bibinfo{pages}{063903},
	\doiprefix\url{10.1103/PhysRevLett.110.063903} (\bibinfo{year}{2013}).
	
	\bibitem{Carminati2015}
	\bibinfo{author}{Carminati, R.} \emph{et~al.}
	\newblock \bibinfo{journal}{\bibinfo{title}{{Electromagnetic density of states
				in complex plasmonic systems}}}.
	\newblock {\emph{\JournalTitle{Surf. Sci. Reps.}}}
	\textbf{\bibinfo{volume}{70}}, \bibinfo{pages}{1--41},
	\doiprefix\url{10.1016/j.surfrep.2014.11.001} (\bibinfo{year}{2015}).
	
	\bibitem{Joulain2005a}
	\bibinfo{author}{Joulain, K.}, \bibinfo{author}{Mulet, J.-P.},
	\bibinfo{author}{Marquier, F.}, \bibinfo{author}{Carminati, R.} \&
	\bibinfo{author}{Greffet, J.-J.}
	\newblock \bibinfo{journal}{\bibinfo{title}{{Surface electromagnetic waves
				thermally excited: Radiative heat transfer, coherence properties and Casimir
				forces revisited in the near field}}}.
	\newblock {\emph{\JournalTitle{Surf. Sci. Reps.}}}
	\textbf{\bibinfo{volume}{57}}, \bibinfo{pages}{59--112},
	\doiprefix\url{10.1016/j.surfrep.2004.12.002} (\bibinfo{year}{2005}).
	
	\bibitem{Castanie2012}
	\bibinfo{author}{Castani{\'{e}}, E.} \emph{et~al.}
	\newblock \bibinfo{journal}{\bibinfo{title}{{Distance dependence of the local
				density of states in the near field of a disordered plasmonic film}}}.
	\newblock {\emph{\JournalTitle{Opt. Lett.}}} \textbf{\bibinfo{volume}{37}},
	\bibinfo{pages}{3006--3008}, \doiprefix\url{10.1364/OL.37.003006}
	(\bibinfo{year}{2012}).
	
	\bibitem{Krachmalnicoff2010}
	\bibinfo{author}{Krachmalnicoff, V.}, \bibinfo{author}{Castani{\'{e}}, E.},
	\bibinfo{author}{{De Wilde}, Y.} \& \bibinfo{author}{Carminati, R.}
	\newblock \bibinfo{journal}{\bibinfo{title}{{Fluctuations of the local density
				of states probe localized surface plasmons on disordered metal films}}}.
	\newblock {\emph{\JournalTitle{Phys. Rev. Lett.}}}
	\textbf{\bibinfo{volume}{105}}, \bibinfo{pages}{183901},
	\doiprefix\url{10.1103/PhysRevLett.105.183901} (\bibinfo{year}{2010}).
	
	\bibitem{Sondergaard2004}
	\bibinfo{author}{S{\o}ndergaard, T.} \& \bibinfo{author}{Bozhevolnyi, S.~I.}
	\newblock \bibinfo{journal}{\bibinfo{title}{{Surface plasmon polariton
				scattering by a small particle placed near a metal surface: An analytical
				study}}}.
	\newblock {\emph{\JournalTitle{Phys. Rev. B}}} \textbf{\bibinfo{volume}{69}},
	\bibinfo{pages}{045422}, \doiprefix\url{10.1103/PhysRevB.69.045422}
	(\bibinfo{year}{2004}).
	
	\bibitem{Kroger1970}
	\bibinfo{author}{Kr{\"{o}}ger, E.} \& \bibinfo{author}{Kretschmann, E.}
	\newblock \bibinfo{journal}{\bibinfo{title}{{Scattering of light by slightly
				rough surfaces or thin films including plasma resonance emission}}}.
	\newblock {\emph{\JournalTitle{Zeit. Phys}}} \textbf{\bibinfo{volume}{237}},
	\bibinfo{pages}{1--15}, \doiprefix\url{10.1007/BF01400471}
	(\bibinfo{year}{1970}).
	
	\bibitem{Bousquet1981}
	\bibinfo{author}{Bousquet, P.}, \bibinfo{author}{Flory, F.} \&
	\bibinfo{author}{Roche, P.}
	\newblock \bibinfo{journal}{\bibinfo{title}{{Scattering from multilayer thin
				films: theory and experiment}}}.
	\newblock {\emph{\JournalTitle{J. Opt. Soc. Am.}}}
	\textbf{\bibinfo{volume}{71}}, \bibinfo{pages}{1115--1123},
	\doiprefix\url{10.1364/JOSA.71.001115} (\bibinfo{year}{1981}).
	
	\bibitem{Gao2013}
	\bibinfo{author}{Gao, Y.} \emph{et~al.}
	\newblock \bibinfo{journal}{\bibinfo{title}{{Plasmonic interferometric sensor
				arrays for high-performance label-free biomolecular detection.}}}
	\newblock {\emph{\JournalTitle{Lab on a Chip}}} \textbf{\bibinfo{volume}{13}},
	\bibinfo{pages}{4755--4764}, \doiprefix\url{10.1039/c3lc50863c}
	(\bibinfo{year}{2013}).
	
	\bibitem{Feng2012a}
	\bibinfo{author}{Feng, J.} \emph{et~al.}
	\newblock \bibinfo{journal}{\bibinfo{title}{{Nanoscale plasmonic
				interferometers for multispectral, high-throughput biochemical sensing}}}.
	\newblock {\emph{\JournalTitle{Nano Lett.}}} \textbf{\bibinfo{volume}{12}},
	\bibinfo{pages}{602--609}, \doiprefix\url{10.1021/nl203325s}
	(\bibinfo{year}{2012}).
	
	\bibitem{Lee2010a}
	\bibinfo{author}{Lee, B.}, \bibinfo{author}{Kim, S.}, \bibinfo{author}{Kim, H.}
	\& \bibinfo{author}{Lim, Y.}
	\newblock \bibinfo{journal}{\bibinfo{title}{{The use of plasmonics in light
				beaming and focusing}}}.
	\newblock {\emph{\JournalTitle{Prog. Quant. Electron.}}}
	\textbf{\bibinfo{volume}{34}}, \bibinfo{pages}{47--87},
	\doiprefix\url{10.1016/j.pquantelec.2009.08.002} (\bibinfo{year}{2010}).
	
	\bibitem{Drezet2008a}
	\bibinfo{author}{Drezet, A.} \emph{et~al.}
	\newblock \bibinfo{journal}{\bibinfo{title}{{Leakage radiation microscopy of
				surface plasmon polaritons}}}.
	\newblock {\emph{\JournalTitle{Mat. Sci. Eng. B}}}
	\textbf{\bibinfo{volume}{149}}, \bibinfo{pages}{220--229},
	\doiprefix\url{10.1016/j.mseb.2007.10.010} (\bibinfo{year}{2008}).
	
	\bibitem{Hohenau2011}
	\bibinfo{author}{Hohenau, A.} \emph{et~al.}
	\newblock \bibinfo{journal}{\bibinfo{title}{{Surface plasmon leakage radiation
				microscopy at the diffraction limit}}}.
	\newblock {\emph{\JournalTitle{Opt. Express.}}} \textbf{\bibinfo{volume}{19}},
	\bibinfo{pages}{25749--25762}, \doiprefix\url{10.1364/OE.19.025749}
	(\bibinfo{year}{2011}).
	
	\bibitem{Stratton1941a}
	\bibinfo{author}{Stratton, J.~A.}
	\newblock \emph{\bibinfo{title}{{Electromagnetic Theory}}}
	(\bibinfo{publisher}{McGraw-Hill, New York}, \bibinfo{year}{1941}).
	
	\bibitem{Morse1953}
	\bibinfo{author}{Morse, P.~M.} \& \bibinfo{author}{Feshbach, H.}
	\newblock \emph{\bibinfo{title}{{Methods of Theoretical Physics}}}
	(\bibinfo{publisher}{McGraw Hill}, \bibinfo{address}{New York},
	\bibinfo{year}{1953}).
	
	\bibitem{Setala2002}
	\bibinfo{author}{Set{\"{a}}l{\"{a}}, T.}, \bibinfo{author}{Shevchenko, A.},
	\bibinfo{author}{Kaivola, M.} \& \bibinfo{author}{Friberg, A.~T.}
	\newblock \bibinfo{journal}{\bibinfo{title}{{Degree of polarization for optical
				near fields}}}.
	\newblock {\emph{\JournalTitle{Phys. Rev. E}}} \textbf{\bibinfo{volume}{66}},
	\bibinfo{pages}{016615}, \doiprefix\url{10.1103/PhysRevE.66.016615}
	(\bibinfo{year}{2002}).
	
	\bibitem{Tervo2003}
	\bibinfo{author}{Tervo, J.}, \bibinfo{author}{Set{\"{a}}l{\"{a}}, T.} \&
	\bibinfo{author}{Friberg, A.~T.}
	\newblock \bibinfo{journal}{\bibinfo{title}{{Degree of coherence for
				electromagnetic fields}}}.
	\newblock {\emph{\JournalTitle{Opt. Express}}} \textbf{\bibinfo{volume}{11}},
	\bibinfo{pages}{1137--1143}, \doiprefix\url{10.1364/OE.11.001137}
	(\bibinfo{year}{2003}).
	
	\bibitem{Volkov2008}
	\bibinfo{author}{Volkov, S.~N.}, \bibinfo{author}{James, D. F.~V.},
	\bibinfo{author}{Shirai, T.} \& \bibinfo{author}{Wolf, E.}
	\newblock \bibinfo{journal}{\bibinfo{title}{{Intensity fluctuations and the
				degree of cross-polarization in stochastic electromagnetic beams}}}.
	\newblock {\emph{\JournalTitle{J. Opt. A: Pure Appl. Opt.}}}
	\textbf{\bibinfo{volume}{10}}, \bibinfo{pages}{055001},
	\doiprefix\url{10.1088/1464-4258/10/5/055001} (\bibinfo{year}{2008}).
	
	\bibitem{Rakic1998}
	\bibinfo{author}{Rakic, A.~D.}, \bibinfo{author}{Djurisic, A.~B.},
	\bibinfo{author}{Elazar, J.~M.} \& \bibinfo{author}{Majewski, M.~L.}
	\newblock \bibinfo{journal}{\bibinfo{title}{{Optical properties of metallic
				films for vertical-cavity optoelectronic devices.}}}
	\newblock {\emph{\JournalTitle{Appl. Opt.}}} \textbf{\bibinfo{volume}{37}},
	\bibinfo{pages}{5271--83}, \doiprefix\url{10.1364/AO.37.005271}
	(\bibinfo{year}{1998}).
	
	\bibitem{Setala2003a}
	\bibinfo{author}{Set{\"{a}}l{\"{a}}, T.}, \bibinfo{author}{Blomstedt, K.},
	\bibinfo{author}{Kaivola, M.} \& \bibinfo{author}{Friberg, A.~T.}
	\newblock \bibinfo{journal}{\bibinfo{title}{{Universality of
				electromagnetic-field correlations within homogeneous and isotropic
				sources}}}.
	\newblock {\emph{\JournalTitle{Phys. Rev. E}}} \textbf{\bibinfo{volume}{67}},
	\bibinfo{pages}{026613}, \doiprefix\url{10.1103/PhysRevE.67.026613}
	(\bibinfo{year}{2003}).
	
	\bibitem{Abramowitz1972}
	\bibinfo{author}{Abramowitz, M.} \& \bibinfo{author}{Stegun, I.~A.}
	\newblock \emph{\bibinfo{title}{{Handbook of Mathematical Functions}}}
	(\bibinfo{publisher}{Dover Publications, New York}, \bibinfo{year}{1972}).
	
\end{thebibliography}
\end{document}


\flushbottom
\maketitle

\thispagestyle{empty}


\section{Derivation of field correlation matrix}
Within the electrostatic approximation the electric field of a surface plasmon polariton (SPP) propagating on a surface defined by the surface profile $\zeta(\boldrho)$, where $\boldrho$  denotes a position vector in the $x$-$y$ plane, is given by $\mathbf{E}(\mathbf{r}) = - \nabla \phi(\mathbf{r})$. The scalar potential $\phi$ is a solution to Laplace's equation $\nabla^2 \phi = 0$ subject to  continuity constraints on $\phi$ and its normal derivative on the supporting interface $\zeta(\boldrho)$ \cite{Farias1983}. The general form of the associated vector Helmholtz equation for the SPP field $\mathbf{E}$ in a source free region can be written in the form 
\begin{align}
\mu(\mathbf{r})\nabla \times \frac{1}{\mu(\mathbf{r})}\nabla \times \mathbf{E}(\mathbf{r}) - \omega^2 \epsilon(\mathbf{r}) \mu(\mathbf{r}) \mathbf{E}(\mathbf{r}) = \mathbf{0}, \label{eq:waveeqn}
\end{align}
where $\mu(\mathbf{r})$ and $\epsilon(\mathbf{r})$ are the spatial distributions of the magnetic permeability and electric permittivity respectively and $\omega$ is the angular frequency of the  wave (assumed monochromatic with an $\exp[-i\omega t]$ time dependence). We henceforth shall consider only non-magnetic media for which $\mu(\mathbf{r}) = \mu_0$ for all $\mathbf{r}$. The electric permittivity distribution describing an arbitrary surface can be written in the form 
\begin{align}
\epsilon(\mathbf{r}) = \epsilon_1 + (\epsilon_2 - \epsilon_1) H(\zeta(\boldrho) - z) \label{eq:rough_epsilon}
\end{align}
where $H(z)$ is the Heaviside function. We also define the permittivity difference $\delta \epsilon(\mathbf{r}) = \epsilon(\mathbf{r}) - {\epsilon}_f(\mathbf{r})$ relative to the permittivity distribution for a flat surface
\begin{align}
\epsilon_f(\mathbf{r}) = \epsilon_1 + (\epsilon_2 - \epsilon_1) H(- z) \label{eq:flat_epsilon}.
\end{align}
Rearranging Eq.~\eqref{eq:waveeqn} thus gives
\begin{align}
\nabla \times \nabla \times \mathbf{E}(\mathbf{r}) - \omega^2 \epsilon_f(\mathbf{r})  \mu_0 \mathbf{E}(\mathbf{r})  = \omega^2 \mu_0 \delta \epsilon(\mathbf{r})\mathbf{E}(\mathbf{r}) \label{eq:waveeqn3}.
\end{align}
The roughness  can thus be modelled as an equivalent source current $\mathbf{J}(\mathbf{r}) = -i \omega \delta \epsilon(\mathbf{r}) \mathbf{E}(\mathbf{r})$.  The general solution for the scattered SPP field can be written directly by employing the Green's tensor $\mathbb{G}(\mathbf{r}_1,\mathbf{r}_2) $ describing elastic scattering of SPPs on a flat surface (defined below). 

If the underlying surface profile is a random field, then so too is the electric field $\mathbf{E}$. Random fields can be partially characterised by their second order moments and we thus consider the two-point field correlation matrix $\mathbb{W}(\mathbf{r}_1,\mathbf{r}_2) = \langle \delta \mathbf{E}^*(\mathbf{r}_1)\delta \mathbf{E}^T(\mathbf{r}_2) \rangle$ here, where $ \delta \mathbf{E}(\mathbf{r}) = \mathbf{E}(\mathbf{r}) - \langle \mathbf{E}(\mathbf{r}) \rangle$. Upon averaging Eq.~\eqref{eq:waveeqn3} we find the mean field $\langle \mathbf{E}(\mathbf{r})\rangle$ satisfies the equation 
\begin{align}
\nabla \times \nabla \times \avg{\mathbf{E}(\mathbf{r}) } - \omega^2 \epsilon_f(\mathbf{r})  \mu_0\avg{\mathbf{E}(\mathbf{r}) } = \omega^2 \mu_0  \avg{\delta\epsilon(\mathbf{r})\mathbf{E}(\mathbf{r})} \label{eq:meanwaveeqn}.
\end{align}
Making a Born approximation, the driving field appearing in $\mathbf{J}$ is replaced by the incident field $\mathbf{E}_0$ which is the solution of Eq.~\eqref{eq:waveeqn} with ${\epsilon} \rightarrow \epsilon_f$, such that $\avg{\delta\epsilon(\mathbf{r})\mathbf{E}(\mathbf{r})} \rightarrow \avg{\delta\epsilon(\mathbf{r})\mathbf{E}_0(\mathbf{r})} = \avg{\delta\epsilon(\mathbf{r})}\avg{\mathbf{E}_0(\mathbf{r})} = \mathbf{0}$ since $ \avg{\delta\epsilon(\mathbf{r})} = 0$ by definition. Hence subtracting Eqs.~\eqref{eq:waveeqn3} and \eqref{eq:meanwaveeqn} we have
\begin{align}
\nabla \times \nabla \times \delta\mathbf{E}(\mathbf{r})  - \omega^2 \epsilon_f(\mathbf{r})  \mu_0 \delta\mathbf{E}(\mathbf{r})  = \omega^2 \mu_0 \delta \epsilon(\mathbf{r})\mathbf{E}_0(\mathbf{r}) \label{eq:waveeqn4}
\end{align}
which has the solution
\begin{align}
\delta\mathbf{E}(\mathbf{r}_1) = i \omega \mu_0 \int \mathbb{G}(\mathbf{r}_1,\mathbf{r}_2) \mathbf{J}(\mathbf{r}_2) d \mathbf{r}_2 \label{eq:Greens_sol},
\end{align}
where we now introduce subscripts on the position vectors to denote different variables.
Using Eq.~\eqref{eq:Greens_sol} it follows that
\begin{align}
\mathbb{W}(\mathbf{r}_1,\mathbf{r}_2)
&= \omega^2 \mu_0^2  \iint \mathbb{G}^*(\mathbf{r}_1,\mathbf{r}_3) \mathbb{W}_J(\mathbf{r}_3,\mathbf{r}_4)    \mathbb{G}^T(\mathbf{r}_2,\mathbf{r}_4)  d \mathbf{r}_3d \mathbf{r}_4 \label{eq:WE_G_WJ_def}
\end{align}
where $\mathbb{W}_J(\mathbf{r}_1,\mathbf{r}_2) = \langle \mathbf{J}^*(\mathbf{r}_1)\mathbf{J}^T(\mathbf{r}_2) \rangle$ is the two point correlation matrix for the current $\mathbf{J}(\mathbf{r}) = -i \omega \delta \epsilon(\mathbf{r}) \mathbf{E}_0(\mathbf{r})$.

In the small roughness regime (i.e $\avg{\zeta(\mathbf{0})^2} \ll \lambda^2$) the permittivity difference can be expanded viz. $\delta{\epsilon}(\mathbf{r})  \approx (\epsilon_2-\epsilon_1)\delta(z)\zeta(\boldrho) + \mathcal{O}(\zeta^2)$, whereby we can write $\mathbf{J}(\mathbf{r}) = \mathbf{j}(\boldsymbol{\rho}) \delta(z)$. We assume that $\mathbf{j}(\boldsymbol{\rho})$ is statistically homogeneous within the plane. With these assumptions  we can write
\begin{align}
\mathbb{W}_J(\mathbf{r}_1,\mathbf{r}_2) = \langle \mathbf{j}(\boldsymbol{\rho}_1)\mathbf{j}^T(\boldsymbol{\rho}_2) \rangle\delta(z_1)\delta(z_2)  = \mathbbold{w}_J(\boldsymbol{\rho}_1,\boldsymbol{\rho}_2) \delta(z_1)\delta(z_2).
\end{align}
Statistical transverse homogeneity of $\mathbf{j}(\boldrho)$ implies that $\mathbb{W}_J(\mathbf{r}_1, \mathbf{r}_1 + \mathbf{P}) = \mathbbold{w}_J(\mathbf{P})  \delta(z_1)\delta(z_2)$ where $\mathbf{P}$ is an arbitrary vector lying in the $x$-$y$ plane. Switching to a Fourier representation will allow simplification of the integrals required to evaluate $\mathbb{W}(\mathbf{r}_1,\mathbf{r}_2) $. Given the symmetry assumed in our system we will consider two-dimensional Fourier transforms. We thus define 
\begin{align}
\widetilde{\mathbf{E}}(\boldsymbol{\kappa},z) &= \int \mathbf{E}(\boldsymbol{\rho},z) e^{-i \boldsymbol{\kappa}\cdot \boldsymbol{\rho}   } d \boldsymbol{\rho}\\
\widetilde{\mathbf{J}}(\boldsymbol{\kappa},z) &= \delta(z)\widetilde{\mathbf{j}}(\boldsymbol{\kappa}) = \delta(z)\int \mathbf{j}(\boldsymbol{\rho}) e^{-i \boldsymbol{\kappa}\cdot \boldsymbol{\rho}   } d \boldsymbol{\rho}.
\end{align}
Consider then 
\begin{align}
\widetilde{\mathbb{W}}_J(\boldsymbol{\kappa}_1,\boldsymbol{\kappa}_2,z_1,z_2) &= \iint \mathbb{W}_J(\mathbf{r}_1,\mathbf{r}_2)  e^{i\boldsymbol{\kappa}_1 \cdot \boldsymbol{\rho}_1 }  e^{-i\boldsymbol{\kappa}_2 \cdot \boldsymbol{\rho}_2 } d\boldsymbol{\rho}_1 d\boldsymbol{\rho}_2 \label{eq:Wtilde_def0}\\
&= \delta(z_1)\delta(z_2)\iint \mathbbold{w}_J(\boldsymbol{\rho}_1,\boldsymbol{\rho}_2)  e^{i\boldsymbol{\kappa}_1 \cdot \boldsymbol{\rho}_1 }e^{-i\boldsymbol{\kappa}_2 \cdot \boldsymbol{\rho}_2 } d\boldsymbol{\rho}_1 d\boldsymbol{\rho}_2\\
&= \delta(z_1)\delta(z_2)\bigg\langle \int  \mathbf{j}^*(\boldsymbol{\rho}_1) e^{i\boldsymbol{\kappa}_1 \cdot \boldsymbol{\rho}_1 }d\boldsymbol{\rho}_1  \int    \mathbf{j}^T(\boldsymbol{\rho}_2)  e^{-i\boldsymbol{\kappa}_2 \cdot \boldsymbol{\rho}_2 } d\boldsymbol{\rho}_2 \bigg\rangle\\
&= \delta(z_1)\delta(z_2)\big\langle \widetilde{\mathbf{j}}^*(\boldsymbol{\kappa}_1)\widetilde{\mathbf{j}}^T(\boldsymbol{\kappa}_2) \big\rangle\\
&= \delta(z_1)\delta(z_2)\widetilde{\mathbbold{w}}_J(\boldsymbol{\kappa}_1,\boldsymbol{\kappa}_2)\label{eq:Wtilde_def}.
\end{align}
We know, however, that for a 2D statistically homogeneous random field that $\mathbbold{w}_J(\boldsymbol{\rho}_1,\boldsymbol{\rho}_2)  = \mathbbold{w}_J(\boldsymbol{\rho}_1-\boldsymbol{\rho}_2) $. Consider then the form of 
\begin{align}
\widetilde{\mathbbold{w}}_J(\boldsymbol{\kappa}_1,\boldsymbol{\kappa}_2) = \iint \mathbbold{w}_J(\boldsymbol{\rho}_1-\boldsymbol{\rho}_2)  e^{i\boldsymbol{\kappa}_1 \cdot \boldsymbol{\rho}_1 }e^{-i\boldsymbol{\kappa}_2 \cdot \boldsymbol{\rho}_2 } d\boldsymbol{\rho}_1 d\boldsymbol{\rho}_2.
\end{align}
Letting $\mathbf{P} = \boldsymbol{\rho}_1-\boldsymbol{\rho}_2$ we can perform a change of variables to give
\begin{align}
\widetilde{\mathbbold{w}}_J(\boldsymbol{\kappa}_1,\boldsymbol{\kappa}_2) &= \iint \mathbbold{w}_J(\mathbf{P})  e^{i\boldsymbol{\kappa}_1 \cdot \boldsymbol{\rho}_1 }e^{-i\boldsymbol{\kappa}_2 \cdot [\boldsymbol{\rho}_1-\mathbf{P}] } d\boldsymbol{\rho}_1 d\mathbf{P}\\
&= \int  e^{i[\boldsymbol{\kappa}_1 -\boldsymbol{\kappa}_2 ]\cdot \boldsymbol{\rho}_1 }   d\boldsymbol{\rho}_1  \int \mathbbold{w}_J(\mathbf{P}) e^{+i\boldsymbol{\kappa}_2 \cdot \mathbf{P} }d\mathbf{P}\\
& = (2\pi)^2 \delta(\boldsymbol{\kappa}_1 - \boldsymbol{\kappa}_2) \widetilde{\mathbbold{w}}_J(-\boldsymbol{\kappa}_2) . \label{eq:wJtilde_form}
\end{align}

The SPP Green's tensor for a planar interface is a function of $\mathbf{P}=\boldsymbol{\rho}_1-\boldsymbol{\rho}_2$ only  \cite{Sondergaard2004,Tomas1995}. We can therefore write the Green's tensor in the form
\begin{align}
\mathbb{G}(\mathbf{r}_1,\mathbf{r}_2) = \mathbb{G}(\boldsymbol{\rho}_1,\boldsymbol{\rho}_2,z_1,z_2) = \mathbb{G}(\mathbf{P},z_1,z_2)
= \frac{1}{(2\pi)^2}\int \widetilde{\mathbb{G}}(\boldsymbol{\kappa},z_1,z_2) e^{i\boldsymbol{\kappa}\cdot \boldsymbol{\rho}} d \boldsymbol{\kappa}.\label{eq:Greens_FT}
\end{align}
The electric field two-point correlation matrix can hence be written using Eqs.~\eqref{eq:WE_G_WJ_def} and \eqref{eq:Greens_FT} as
\begin{align}
&\mathbb{W}(\mathbf{r}_1,\mathbf{r}_2) =  \frac{\omega^2\mu_0^2}{(2\pi)^4} \iint d \mathbf{r}_3d \mathbf{r}_4 
\left[\int \widetilde{\mathbb{G}}^*(\boldsymbol{\kappa}_1,z_1,z_3) e^{-i\boldsymbol{\kappa}_1\cdot [\boldsymbol{\rho}_1 -\boldsymbol{\rho}_3 ]  } d \boldsymbol{\kappa}_1   \right]
\mathbb{W}_J(\mathbf{r}_3,\mathbf{r}_4)  
\left[ \int \widetilde{\mathbb{G}}^T(\boldsymbol{\kappa}_2,z_2,z_4) e^{i\boldsymbol{\kappa}_2\cdot [\boldsymbol{\rho}_2-\boldsymbol{\rho}_4 ] } d \boldsymbol{\kappa}_2\right].
\end{align}
Rearranging and reordering the integration gives
\begin{align}
&\mathbb{W}(\mathbf{r}_1,\mathbf{r}_2) =  \frac{\omega^2\mu_0^2}{(2\pi)^4}\iint d \boldsymbol{\kappa}_1 d \boldsymbol{\kappa}_2  e^{-i\boldsymbol{\kappa}_1\cdot \boldsymbol{\rho}_1   } e^{i\boldsymbol{\kappa}_2\cdot \boldsymbol{\rho}_2 } 
\widetilde{\mathbb{G}}^*(\boldsymbol{\kappa}_1,z_1,z_3)\left[\iint\mathbb{W}_J(\mathbf{r}_3,\mathbf{r}_4) e^{i\boldsymbol{\kappa}_1\cdot \boldsymbol{\rho}_3  } e^{-i\boldsymbol{\kappa}_2\cdot \boldsymbol{\rho}_4  } d \mathbf{r}_3d \mathbf{r}_4\right]
\widetilde{\mathbb{G}}^T(\boldsymbol{\kappa}_2,z_2,z_4) .
\end{align}
Further using Eqs.~\eqref{eq:Wtilde_def0}, \eqref{eq:Wtilde_def} and \eqref{eq:wJtilde_form} this expression reduces to
\begin{align}
&\mathbb{W}(\mathbf{r}_1,\mathbf{r}_2) =  \frac{\omega^2\mu_0^2}{(2\pi)^2} \iint d \boldsymbol{\kappa}_1 d \boldsymbol{\kappa}_2  e^{-i\boldsymbol{\kappa}_1\cdot \boldsymbol{\rho}_1   } e^{i\boldsymbol{\kappa}_2\cdot \boldsymbol{\rho}_2 }   \delta(\boldsymbol{\kappa}_1 - \boldsymbol{\kappa}_2)
\iint dz_3dz_4 \widetilde{\mathbb{G}}^*(\boldsymbol{\kappa}_1,z_1,z_3)
\delta(z_3)\delta(z_4) \widetilde{\mathbbold{w}}_J(-\boldsymbol{\kappa}_1)
\widetilde{\mathbb{G}}^T(\boldsymbol{\kappa}_2,z_2,z_4) .
\end{align}
The Dirac delta functions allow the integration of $z_3,z_4$ and $\boldsymbol{\kappa}_2$ to be easily performed, yielding
\begin{align}
\mathbb{W}(\mathbf{r}_1,\mathbf{r}_2) &=  \frac{\omega^2\mu_0^2}{(2\pi)^2}\int d \boldsymbol{\kappa} \, e^{-i\boldsymbol{\kappa}\cdot\mathbf{P}  }  \widetilde{\mathbb{G}}^*(\boldsymbol{\kappa},z_1,0)
\widetilde{\mathbbold{w}}_J(-\boldsymbol{\kappa})
\widetilde{\mathbb{G}}^T(\boldsymbol{\kappa},z_2,0) \label{eq:Greens_Fourier}\\
& =  \frac{\omega^2\mu_0^2}{(2\pi)^2}\iint d \boldsymbol{\kappa}  \,d\mathbf{Q} \,e^{-i\boldsymbol{\kappa}\cdot(\mathbf{P}-\mathbf{Q})  }  \widetilde{\mathbb{G}}^*(\boldsymbol{\kappa},z_1,0)
\mathbbold{w}_J(\mathbf{Q}) 
\widetilde{\mathbb{G}}^T(\boldsymbol{\kappa},z_2,0) \label{eq:Greens_Fourier2},
\end{align}
where we have also dropped the unneeded subscript on $\boldsymbol{\kappa}_1$ and used the definition of $\widetilde{\mathbbold{w}}_J(-\boldsymbol{\kappa})$ in the final equality.

The SPP contribution to the total Green's tensor for a single planar interface is given by Eq. (21) of \cite{Sondergaard2004}. Specifically assuming an interface at $z=0$ separates a dielectric medium ($z>0$) with permittivity $\epsilon_1$ and a metallic medium ($z<0$) of complex permittivity $\epsilon_2$ ($\text{Im}[\epsilon_2]>0$) the SPP Green's tensor can be shown to take the form 
\begin{align}
&\mathbb{G}(\mathbf{r}_1,\mathbf{r}_2) = \int_0^{\infty} \frac{\kappa^2}{K_0 (k_{\txtpow{SPP}}^2 - \kappa^2 )} d\kappa \int_0^{2\pi} d\phi \Big[\hat{\mathbf{z}} - i \alpha \hat{\boldsymbol{\kappa}}\Big]\Big[\hat{\mathbf{z}}^T  + i \alpha \hat{\boldsymbol{\kappa}}^T\Big] e^{i\boldsymbol{\kappa} \cdot \mathbf{P}} e^{-\alpha\kappa (z_1+z_2)}\label{eq:GSPP}
\end{align}
for $z_1,z_2\geq 0$, where $\alpha = \sqrt{-\epsilon_1/\epsilon_2}$, $\boldsymbol{\kappa} = \kappa[\cos\phi,\sin\phi,0]^T$,  $\hat{\boldsymbol{\kappa}} = \boldsymbol{\kappa}  / \kappa$, 
\begin{align}
K_0 &= 2\pi^2 \sqrt{-\epsilon_1\epsilon_2} \left(1-\frac{\epsilon_1^2}{\epsilon_2^2}\right) \frac{\epsilon_1 +\epsilon_2} {\epsilon_1\epsilon_2},
\end{align}
$k_{\txtpow{SPP}} = k_0 \sqrt{{\epsilon_1\epsilon_2}/({\epsilon_1 +\epsilon_2})}$, $k_0 = \omega / c$ and   $c$ is the speed of light in vacuum.  Comparison with Eq.~\eqref{eq:Greens_FT} yields
\begin{align}
\widetilde{\mathbb{G}}(\boldsymbol{\kappa},z_1,z_2) &= \frac{(2\pi)^2\kappa}{K_0 (k_{\txtpow{SPP}}^2 - \kappa^2 )} e^{-\alpha\kappa (z_1+z_2)}\Big[\hat{\mathbf{z}} - i \alpha \hat{\boldsymbol{\kappa}}\Big]\Big[\hat{\mathbf{z}}^T  + i \alpha \hat{\boldsymbol{\kappa}}^T\Big] =  g(\kappa,z_1,z_2)\mathbb{M}(\phi) \label{eq:GSPPtilde}
\end{align}
where $g(\kappa,z_1,z_2) = {(2\pi)^2\kappa}e^{-\alpha\kappa (z_1+z_2)} / [{K_0(k_{\txtpow{SPP}}^2 - \kappa^2)} ]$ and 
\begin{align}
\mathbb{M}(\phi) &= \Big[\hat{\mathbf{z}} - i \alpha \hat{\boldsymbol{\kappa}}\Big]\Big[\hat{\mathbf{z}}^T  + i \alpha \hat{\boldsymbol{\kappa}}^T\Big] = \hat{\mathbf{z}}\hat{\mathbf{z}}^T + i \alpha[\hat{\mathbf{z}} \hat{\boldsymbol{\kappa}}^T + \hat{\boldsymbol{\kappa}}\hat{\mathbf{z}}^T] - \alpha^2 \hat{\boldsymbol{\kappa}}\hat{\boldsymbol{\kappa}}^T\nonumber\\
&= \left[  \begin{array}{ccc}
\frac{1}{2}\alpha^2 (1+ \cos 2 \phi) & \frac{1}{2}\alpha^2\sin 2\phi & -i\alpha\cos\phi \\
\frac{1}{2}\alpha^2 \sin 2\phi & \frac{1}{2}\alpha^2 (1-\cos 2 \phi) & -i\alpha\sin\phi \\ 
i\alpha \cos\phi & i\alpha \sin\phi & 1 
\end{array}\right].
\end{align}
Expressing $\mathbbold{w}_J(\mathbf{Q})$ in the form
\begin{align}
\mathbbold{w}_J(\mathbf{Q}) = \sum_{j=1}^3\sum_{k=1}^3 w_{jk}(\mathbf{Q}) \hat{\mathbf{e}}_j(\beta) \hat{\mathbf{e}}_k^T(\beta)
\end{align}
where $\mathbf{Q} = Q[\cos\beta,\sin\beta,0]^T$ and
\begin{align}
\hat{\mathbf{e}}_1(\beta) &= \hat{\mathbf{Q}} = \cos\beta \hat{\mathbf{x}} + \sin\beta\hat{\mathbf{y}}\\
\hat{\mathbf{e}}_2(\beta) &= \hat{\boldsymbol{\beta}} = -\sin\beta \hat{\mathbf{x}} + \cos\beta\hat{\mathbf{y}}\\
\hat{\mathbf{e}}_3(\beta) &=\hat{\mathbf{z}},
\end{align}
are unit vectors in the radial, azimuthal and longitudinal directions respectively,  we can use Eq.~\eqref{eq:GSPPtilde} to give
\begin{align}
\mathbb{W}(\mathbf{r}_1,\mathbf{r}_2)&=   \frac{(2\pi)^2\omega^2\mu_0^2}{|K_0|^2} \sum_{j,k = 1}^3 \int  w_{jk}(\mathbf{Q}) \mathbb{I}_{jk}(\mathbf{P}-\mathbf{Q})  d\mathbf{Q},\label{eq:WSPPint1}
\end{align} 
where letting $\mathcal{Z} = \alpha^* z_1 + \alpha z_2$
\begin{align}
&\mathbb{I}_{jk}(\mathbf{P}-\mathbf{Q},z_1,z_2)  =\int_0^\infty \int_0^{2\pi} \frac{\kappa^3 e^{-i\boldsymbol{\kappa}\cdot(\mathbf{P}-\mathbf{Q})  } e^{-\kappa \mathcal{Z}}  d \kappa d\phi}{(k_{\txtpow{SPP}}^2 - \kappa^2)(k_{\txtpow{SPP}}^2 - \kappa^2)^*} \mathbb{M}^*(\phi) \hat{\mathbf{e}}_j(\beta) \hat{\mathbf{e}}_k^T(\beta) \mathbb{M}^T(\phi)  
\label{eq:WSPPint2}.
\end{align}
The matrix factor can be evaluated and takes the form
\begin{align}
\mathbb{M}_{jk}(\mathbf{P}-\mathbf{Q}) &= \mathbb{M}^*(\phi) \hat{\mathbf{e}}_j(\beta) \hat{\mathbf{e}}_k^T(\beta) \mathbb{M}^T(\phi) =  m_{jk} \mathcal{M}_{jk}(\beta,\phi) \mathbb{M}_{33}(\phi) \label{eq:MijMAT}
\end{align}
where
\begin{align}
\mathbb{M}_{33}(\phi) = \left[\begin{array}{ccc}
\frac{1}{2}|\alpha|^2 (1+\cos 2\phi) & \frac{1}{2}|\alpha|^2 \sin 2\phi & i \alpha^* \cos\phi \\
\frac{1}{2}|\alpha|^2 \sin 2\phi & \frac{1}{2}|\alpha|^2 (1-\cos 2 \phi) & i \alpha^* \sin\phi \\
-i\alpha \cos\phi & -i\alpha \sin \phi & 1
\end{array}\right]
\end{align}
$m_{11} = m_{22} =\frac{1}{2}|\alpha|^2$, $m_{12} = m_{21}^* = -\frac{1}{2}|\alpha|^2$, $m_{13} = m_{31}^* = -i\alpha^*$, $m_{23} = m_{32}^* = i\alpha^*$, $m_{33} =1$ and 
\begin{align}
\mathcal{M}_{11}(\beta,\phi) &= 1 + \cos 2(\beta-\phi) \\ 
\mathcal{M}_{22}(\beta,\phi) &= 1 - \cos 2(\beta-\phi) \\ 
\mathcal{M}_{33}(\beta,\phi) &= 1 \\ 
\mathcal{M}_{12}(\beta,\phi) &=\mathcal{M}_{21}(\beta,\phi) = \sin 2(\beta-\phi)\\
\mathcal{M}_{13}(\beta,\phi) &=\mathcal{M}_{31}(\beta,\phi) = \cos (\beta-\phi)\\
\mathcal{M}_{23}(\beta,\phi) &=\mathcal{M}_{32}(\beta,\phi) = \sin (\beta-\phi).
\end{align}

At this point we make the further restriction that $w_{jk}(\mathbf{Q}) = w_{jk}(Q)$, i.e. the two-point current correlation matrix is a function of the separation of the two points only and not their relative direction.  For this case we have
\begin{align}
&\mathbb{W}(\mathbf{r}_1,\mathbf{r}_2)=\frac{(2\pi)^2\omega^2\mu_0^2}{|K_0|^2} \sum_{j,k = 1}^3 m_{jk} \int_0^\infty Q\,  w_{jk}({Q}) \mathbb{I}_{jk}^\beta(\mathbf{P}-\mathbf{Q},z_1,z_2)  d{Q},\label{eq:WSPPint1}
\end{align} 
where using Eq.~\eqref{eq:MijMAT}, letting $\mathbf{P} = P[\cos\theta,\sin\theta,0]^T$ and splitting the exponential term we can write
\begin{align}
&\mathbb{I}_{jk}^\beta (\mathbf{P}-\mathbf{Q},z_1,z_2)  = \int_0^\infty \frac{\kappa^3 e^{-\kappa \mathcal{Z}}  }{(k_{\txtpow{SPP}}^2 - \kappa^2)(k_{\txtpow{SPP}}^2 - \kappa^2)^*} \int_0^{2\pi} \!\! e^{-i{\kappa}{P} \cos(\theta - \phi) } \mathbb{M}_{33}(\phi)
\left[\int_0^{2\pi}\!\!\! \mathcal{M}_{jk}(\beta,\phi)e^{i{\kappa}{Q} \cos(\beta - \phi) } d\beta\right]\! d\phi d \kappa \label{eq:Iijbetaint}.
\end{align} 
Recalling the integration identity \cite{Stratton1941a}
\begin{align}
\int_0^{2\pi} \left\{\begin{array}{c}\cos n \beta \\ \sin n \beta\end{array}\right\} e^{-ia\cos(\beta -\gamma)} d\beta = 2\pi (-i)^n\left\{\begin{array}{c}\cos n \gamma \\ \sin n \gamma\end{array}\right\}  J_n(a).\label{eq:Besselint}
\end{align}
the angular integration over $\beta$ in Eq.~\eqref{eq:Iijbetaint} can be performed yielding
\begin{align}
2\pi B_{jk} &= \int_0^{2\pi} \mathcal{M}_{jk}(\beta,\phi)e^{i{\kappa}{Q} \cos(\beta - \phi) } d\beta\nonumber\\
& = \left\{\begin{array}{lll} 
2\pi[J_0(\kappa Q) - J_2(\kappa Q)] &\text{for~} (j,k) = (1,1)\\
2\pi[J_0(\kappa Q) + J_2(\kappa Q)] &\text{for~} (j,k) = (2,2)\\
2\pi J_0(\kappa Q) &\text{for} (j,k) = (3,3)\\
2\pi i J_1(\kappa Q)  &\text{for~} (j,k) = (1,3) \text{~and~} (3,1)\\
0 & \text{otherwise}
\end{array} \right.
\end{align}
Noting that these expressions are independent of $\phi$ we can similarly evaluate the integral over $\phi$ yielding
\begin{align}
&\mathbb{I}_{jk}^\beta (\mathbf{P}-\mathbf{Q},z_1,z_2)  = (2\pi)^2 \int_0^\infty \frac{\kappa^3{B}_{jk}(\kappa Q) e^{-\kappa \mathcal{Z}}  }{(k_{\txtpow{SPP}}^2 - \kappa^2)(k_{\txtpow{SPP}}^2 - \kappa^2)^*}  \mathbb{N}(\kappa P,\theta)  d\kappa
\end{align}
where

	\begin{align}
	\mathbb{N}(\kappa P,\theta) = \left[\begin{array}{ccc}  
	\frac{1}{2}|\alpha|^2 [J_0(\kappa P) - J_2(\kappa P) \cos 2 \theta]& - \frac{1}{2} |\alpha|^2 J_2(\kappa P) \sin 2 \theta & \alpha^* J_1(\kappa P) \cos \theta \\
	- \frac{1}{2} |\alpha|^2 J_2(\kappa P) \sin 2 \theta  &\frac{1}{2}|\alpha|^2 [J_0(\kappa P) + J_2(\kappa P) \cos 2 \theta]& \alpha^* J_1(\kappa P) \sin \theta \\
	-\alpha J_1(\kappa P) \cos \theta & - \alpha J_1(\kappa P)  \sin \theta & J_0(\kappa P) 
	\end{array}\right].
	\end{align}
Now defining the auxiliary matrix
\begin{align}
\mathbb{V}_n(P,Q,\theta,z_1,z_2) = \int_0^\infty \frac{\kappa^3 J_n(\kappa Q) e^{-\kappa \mathcal{Z}}  }{(k_{\txtpow{SPP}}^2 - \kappa^2)(k_{\txtpow{SPP}}^2 - \kappa^2)^*}  \mathbb{N}(\kappa P,\theta)  d\kappa
\end{align} 
we have 
\begin{align}
\mathbb{I}_{jk}^\beta (\mathbf{P}-\mathbf{Q})  = 
 (2\pi)^2  [\delta_{jk} \{\mathbb{V}_0 + (\delta_{j2} - \delta_{j1}) \mathbb{V}_2 \}  +  i(\delta_{j1}\delta_{k3} + \delta_{j3}\delta_{k1}) \mathbb{V}_1]
\end{align}
and hence
\begin{align}
\mathbb{W}(\mathbf{r}_1,\mathbf{r}_2)=   \frac{(2\pi)^4\omega^2\mu_0^2}{|K_0|^2} \sum_{j,k = 1}^3 m_{jk} &\int_0^\infty  \bigg[\delta_{jk} \{\mathbb{V}_0(P,Q,\theta,z_1,z_2) + (\delta_{j2} - \delta_{j1}) \mathbb{V}_2(P,Q,\theta,z_1,z_2) \} \nonumber\\
& \quad \quad \quad \quad  +  i(\delta_{j1}\delta_{k3} + \delta_{j3}\delta_{k1}) \mathbb{V}_1(P,Q,\theta,z_1,z_2)\bigg] Q\,  w_{jk}({Q})d{Q}\label{eq:WSPPfinal0}.
\end{align} 
Eq.~\eqref{eq:WSPPfinal0} can however be simplified further by considering the assumption $w_{jk}(\mathbf{Q}) = w_{jk}(Q)$ further. In particular, we require this angular independence to hold regardless of the two points $\mathbf{r}_1$ and $\mathbf{r}_2$ under consideration, i.e. for general $\mathbf{Q}$. Satisfying this requirement, necessitates that $\mathbbold{w}_J$ must be invariant under rotations around the $z$ axis, and therefore $w_{11}=w_{22}$, $w_{12}=-w_{21}$, $w_{13} = w_{31} = w_{23} = w_{32} = 0$. Rewriting Eq.~\eqref{eq:WSPPfinal0} in the form
\begin{align}
\mathbb{W}(\mathbf{P},z_1,z_2)=   \frac{(2\pi)^4\omega^2\mu_0^2}{|K_0|^2}  \sum_{k=0}^2  & \int_0^\infty c_{k}({Q}) \mathbb{V}_{k}(P,Q,\theta,z_1,z_2) dQ
\end{align}
where, recalling $m_{11} = m_{22}$,
\begin{align}
c_{0}(Q) &= \sum_{j,k = 1}^3m_{jk}Q\,\delta_{ij}w_{jk}(Q) \equiv Q\{|\alpha|^2  w_{11}(Q) + w_{33}(Q) \}\\
c_{1}(Q) &= i\sum_{j,k = 1}^3 m_{jk}Q\,(\delta_{j1}\delta_{k3} + \delta_{j3}\delta_{k1})w_{jk}(Q) \equiv 0\\
c_{2}(Q) &= \sum_{j,k = 1}^3m_{jk}Q\, \delta_{jk}(\delta_{j2} - \delta_{j1})w_{jk}(Q) \equiv 0
\end{align}
and  $\delta_{jk}$ is the Kronecker delta, we find the simplified form for the correlation matrix as 
\begin{align}
&\mathbb{W}(\mathbf{P},z_1,z_2)= \frac{(2\pi)^4\omega^2\mu_0^2 }{|K_0|^2}   \int_0^\infty  Q \{|\alpha|^2  w_{11}(Q) + w_{33}(Q) \}\mathbb{V}_{0}(P,Q,\theta,z_1,z_2) dQ\label{eq:WSPPfinal}.
\end{align}

\begin{figure}[t!]
	\begin{center}
		\includegraphics[width=0.5\columnwidth]{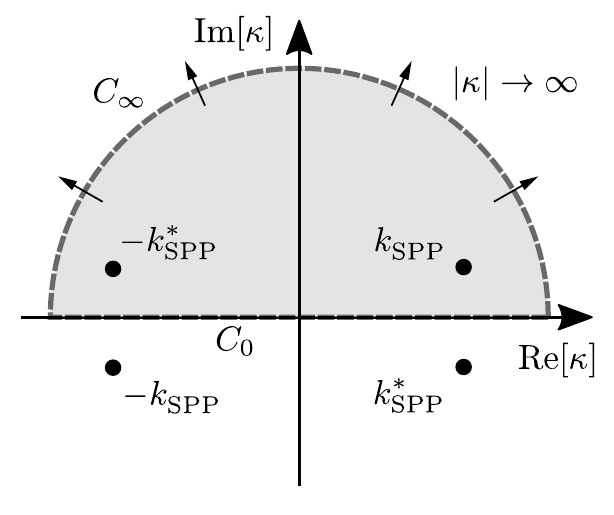}
		\caption{Integration contour on the complex $\kappa$ plane used to evaluate $L_{mn}$ and positions of the SPP related poles. }\label{fig:intcontour}
	\end{center}
\end{figure}

Our final task thus remains to find an expression for $\mathbb{V}_0$ (we can also find expressions for $\mathbb{V}_n$ without great effort and thus we keep things general for the sake of interest). In particular we must evaluate the integrals
\begin{align}
L_{nm} = \int_0 ^\infty  \frac{\kappa^3 J_n(\kappa Q) J_m(\kappa P)  e^{-\kappa \mathcal{Z}}  }{(k_{\txtpow{SPP}}^2 - \kappa^2)(k_{\txtpow{SPP}}^2 - \kappa^2)^*}  d\kappa
\end{align}
which make up the individual terms in $\mathbb{V}_n$ viz.
\begin{align}
	\mathbb{V}_n(P,Q,\theta,z_1,z_2) = \left[\begin{array}{ccc} 
	\frac{1}{2}|\alpha|^2 [L_{n0}(P,Q) - L_{n2}(P,Q) \cos 2 \theta]& - \frac{1}{2} |\alpha|^2 L_{n2}(P,Q) \sin 2 \theta & \alpha^* L_{n1}(P,Q) \cos \theta \\
	- \frac{1}{2} |\alpha|^2 L_{n2}(P,Q) \sin 2 \theta  &\frac{1}{2}|\alpha|^2 [L_{n0}(P,Q) + L_{n2}(P,Q) \cos 2 \theta]& \alpha^* L_{n1}(P,Q) \sin \theta \\
	-\alpha L_{n1}(P,Q) \cos \theta & - \alpha L_{n1}(P,Q)  \sin \theta & L_{n0}(P,Q) 
	\end{array}\right]. \label{eq:Vn}
\end{align}
We begin by first noting that $J_m(\kappa P) = [H_m^{(1)}(\kappa P) + H_m^{(2)}(\kappa P)]/2$ where $H_m^{(1,2)}$ are the Hankel functions of the first and second kind of order $m$. This substitution however introduces a pole at $\kappa= 0$ that must be excluded, such that we consider
\begin{align}
L_{nm} = \frac{1}{2}\lim_{\delta \rightarrow 0 }&\left[\int_{\delta}^{\infty}\frac{\kappa^3 J_n(\kappa Q)  H_m^{(1)}(\kappa P) e^{-\kappa \mathcal{Z}}  }{(k_{\txtpow{SPP}}^2 - \kappa^2)(k_{\txtpow{SPP}}^2 - \kappa^2)^*}  d\kappa  + \int_{\delta}^{\infty}\frac{\kappa^3 J_n(\kappa Q)  H_m^{(2)}(\kappa P)e^{-\kappa \mathcal{Z}}  }{(k_{\txtpow{SPP}}^2 - \kappa^2)(k_{\txtpow{SPP}}^2 - \kappa^2)^*}  d\kappa \right]. \label{eq:Lnm_Hankel}
\end{align}
Now letting $\kappa = e^{-i\pi} \kappa'$ and using the reflection formulae $H_m^{(2)}(e^{-i\pi} z) = -e^{-m \pi i} H_m^{(1)}(z)$ and $J_n(e^{-i\pi} z) = e^{-n \pi i} J_n(z)$ \cite{Abramowitz1972}  we have
\begin{align}
L_{nm} &= \frac{1}{2}\lim_{\delta \rightarrow 0 }\left[\int_{\delta}^{\infty}\frac{\kappa^3 J_n(\kappa Q) H_m^{(1)}(\kappa P)  e^{-\kappa \mathcal{Z}}  }{(k_{\txtpow{SPP}}^2 - \kappa^2)(k_{\txtpow{SPP}}^2 - \kappa^2)^*}   d\kappa + (-1)^{n+m}\int_{-\infty}^{-\delta}\frac{\kappa'^3 J_n(\kappa' Q) H_m^{(1)}(\kappa' P)  e^{\kappa' \mathcal{Z}}  }{(k_{\txtpow{SPP}}^2 - \kappa'^2)(k_{\txtpow{SPP}}^2 - \kappa'^2)^*}  d\kappa' \right].
\end{align}
The two integrals can be combined into a single integral according to
\begin{align}
\!\!L_{nm} = \frac{1}{2} \text{PV}\left[\int_{-\infty}^{\infty}g(\kappa)  d\kappa \right] = \frac{1}{2} \left[\int_{C_0} g(\kappa)  d\kappa - \underset{\kappa=0}{\text{Res}}[g(\kappa)] \right] \label{eq:Lnm_int_cont}
\end{align}
where PV denotes the principal value, Res denotes the complex residue, the integration contour $C_0$ is shown in Figure~\ref{fig:intcontour} and $g(\kappa)$ is an analytic function in the upper half plane given by
\begin{align}
g(\kappa) = (\text{sign}[\text{Re}(\kappa)])^{n+m}\frac{\kappa^3 J_n(\kappa Q) H_m^{(1)}(\kappa P)   }{(k_{\txtpow{SPP}}^2 - \kappa^2)(k_{\txtpow{SPP}}^2 - \kappa^2)^*} \exp[-\kappa\sigma(\kappa) \mathcal{Z}]
\end{align}
with $\sigma = \mbox{sign}[\mbox{Re}(\kappa)]$.
To examine the residue at $\kappa = 0$ we can consider the small $\kappa$ behaviour of $g(\kappa)$ using the small argument expansions \cite{Abramowitz1972}
\begin{align}
J_n(z) &\sim \left(\frac{z}{2}\right)^n / \Gamma(n+1) \\
H_n^{(1)}(z) &\sim -\frac{i}{\pi} \Gamma(n) \left(\frac{z}{2}\right)^{-n},
\end{align}
where $\Gamma(n)$ is the Gamma function.
Specifically
\begin{align}
\underset{\kappa=0}{\text{Res}}[g(\kappa)] &= \lim_{\kappa\rightarrow 0} [\kappa 
g(\kappa)]= \lim_{\kappa\rightarrow 0} \left[\kappa^{n-m} \left(\frac{\kappa^4}{|k_{\txtpow{SPP}}|^4} + \ldots \right)\right]
\end{align}
which tends to zero for $m-n\geq 3 $ cases considered in this work.

With a view to extending the integration contour $C_0$ we consider the behaviour of $J_n(\kappa Q)H_m^{(1)}(\kappa P)$ for $|\kappa| \rightarrow \infty$. The asymptotic formulae
\begin{align}
J_n(z)&=\sqrt{\frac{2}{\pi z}} \cos(z - n\pi/2 - \pi/4) + \cdots \\
H_n^{(1)}(z) &= \sqrt{\frac{2}{\pi z}} \exp[i(z - n\pi/2 - \pi/4)]+ \cdots
\end{align}
give 
\begin{align}
J_n(\kappa Q)H_m^{(1)}(\kappa P) \rightarrow \frac{2}{\pi\kappa \sqrt{PQ}} &\left[e^{i\kappa(P+Q)} e^{-i\pi (n+m+1)/2} + e^{i \kappa(P-Q)}e^{i\pi(n-m)/2}\right].
\end{align}
Splitting $\kappa = \kappa_r + i \kappa_i$ into its real and imaginary parts, we see that  the two position dependent exponents scale as $e^{i\kappa (P\pm Q)} \sim e^{-\kappa_i(P\pm Q)}$, both of which decay for $\kappa_i >0 $, $|\kappa| \rightarrow \infty$ if $P>Q$. Assuming this latter condition to hold true, the integration contour $C_0$ in Eq.~\eqref{eq:Lnm_int_cont} can be extended to include the semicircular contour $C_\infty$ (as shown in Figure~\ref{fig:intcontour}) where the radius of $C_\infty$ is taken to infinity, since the value of the integrand, $g(\kappa)$, is zero along this contour. Accordingly the integration path is now a closed loop and noting Jordan's lemma, the integral can be evaluated using the residual theorem. The function $g(\kappa)$ has four simple poles at $\kappa = \pm k_{\txtpow{SPP}}$ and $\pm k^*_{\txtpow{SPP}}$, only two of which lie within the integration contour namely  $\kappa_1 = k_{\txtpow{SPP}}$ and $\kappa_2 = -k_{\txtpow{SPP}}^*$. Accordingly
\begin{align}
L_{nm} &= \pi i \left[ \underset{\kappa = k_{\mbox{\tiny{SPP}}}}{\text{Res}}[g(\kappa)]  +  \underset{\kappa = -k^*_{\mbox{\tiny{SPP}}}}{\text{Res}}[g(\kappa)]  \right] \\
&= \pi i\left[ \lim_{\kappa \rightarrow k_{\mbox{\tiny{SPP}}}} \left[(\kappa-k_{\mbox{\tiny{SPP}}}) g(\kappa)\right]+\lim_{\kappa \rightarrow -k_{\mbox{\tiny{SPP}}}^*} \left[(\kappa+k_{\mbox{\tiny{SPP}}}^*) g(\kappa)\right]\right].
\end{align}
Evaluating the limits gives $L_{nm}(P,Q) = l_{nm}(P,Q,k_{\txtpow{SPP}})  +l_{nm}(P,Q,-k^*_{\txtpow{SPP}}) $
where
\begin{align}
l_{nm}(P,Q,\kappa) = \sigma(\kappa)^{n+m}\frac{\pi}{4}\frac{\kappa^2 } {\mbox{Im}[\kappa^2] } J_n(\kappa Q)H_m^{(1)}(\kappa P)\exp[-\kappa\sigma(\kappa) \mathcal{Z}].
\end{align}
For the case $P < Q$, instead of initially expanding $J_m(\kappa P)$ in terms of the Hankel functions in Eq.~\eqref{eq:Lnm_Hankel}, we make the alternative substitution $J_n(\kappa Q) = [H_n^{(1)}(\kappa Q) + H_n^{(2)}(\kappa Q)]/2$ which upon following the same logic as above gives
\begin{align}
l_{nm}(P,Q,\kappa ) = \Lambda_{nm}(\kappa)
\left\{  \begin{array}{ll} J_n(\kappa Q)H_m^{(1)}(\kappa P) &\text{for~} P > Q \\
H_n^{(1)}(\kappa Q ) J_m(\kappa P) & \text{for~} P < Q
\end{array}\right. \label{eq:Lmn}
\end{align}
where $\Lambda_{nm}(\kappa) = \sigma(\kappa)^{n+m}{\pi}{\kappa^2 }\exp[-\kappa\sigma(\kappa) \mathcal{Z}]/ (4 {\mbox{Im}[\kappa^2] })$.

\section{Degree of polarisation}
In this section we evaluate the 2D and 3D degrees of polarisation (DOP) at $\mathbf{P} = \mathbf{0}$ defined by \cite{Setala2002,Tervo2003}
\begin{align}
(\mathcal{D}_{2\txtpow{D}})^2 = 2\left\{ \frac{\|\mathbb{W}(\mathbf{0},z_1,z_2)\|_F^2}{\text{tr}[\mathbb{W}(\mathbf{0},z_1,z_2)]^2} -\frac{1}{2} \right\}, \label{eq:DOP2}
\end{align}
and 
\begin{align}
(\mathcal{D}_{3\txtpow{D}})^2 = \frac{3}{2}\left\{ \frac{\|\mathbb{W}^2(\mathbf{0},z_1,z_2)\|_F^2}{\text{tr}[\mathbb{W}(\mathbf{0},z_1,z_2)]^2} -\frac{1}{3} \right\}, \label{eq:DOP3}
\end{align}
respectively where $\|\cdots \|_F$ denotes the Frobenius norm. We thus must first evaluate $\mathbb{W}(\mathbf{0},z_1,z_2)$. Using Eqs.~\eqref{eq:Vn} and \eqref{eq:Lmn} at $P=0$ we have $l_{nm}(0,Q,\kappa) =  \Lambda_{nm}(\kappa) H_n^{(1)}(\kappa Q) \delta_{m0} $  such that
\begin{align}
\mathbb{V}_n(0,Q,\theta,z_1,z_2) = \left[\begin{array}{ccc} 
\frac{1}{2}|\alpha|^2 &0&0 \\
0 & \frac{1}{2}|\alpha|^2 & 0 \\
0 & 0 & 1\end{array}\right] L_{n0} \triangleq \mathbb{A} L_{n0}
\end{align}
and 
\begin{align}
&\mathbb{W}(\mathbf{P} = \mathbf{0},z_1,z_2)=   \mathbb{A}\Bigg[ \frac{(2\pi)^4\omega^2\mu_0^2 |\alpha|^2}{|K_0|^2}  \int_0^\infty Q\, w_{11}({Q})  L_{00}(0,Q)   d{Q} \Bigg]= W_0  \mathbb{A}\label{eq:WSPP_Pzero2},
\end{align} 
where $W_0$ is a constant found by evaluating the integral. Substituting this expression into Eq.~\eqref{eq:DOP3} directly gives 
\begin{align}
\mathcal{D}_{\txtpow{3D}} = \frac{|\alpha|^2 -2}{2(1+|\alpha|^2 )}.
\end{align}
For the 2D case describing correlations between in-plane field components we consider the reduced correlation matrix 
\begin{align}
\mathbb{W}_\parallel(\mathbf{P}=0,z_1,z_2) = W_0   \left[\begin{array}{cc} 
\frac{1}{2}|\alpha|^2 &0 \\
0 & \frac{1}{2}|\alpha|^2\end{array}\right] 
\end{align}
yielding $\mathcal{D}_{2\txtpow{D}}^\parallel = 0$, whilst to describe the 2D DOP between in and out-of-plane field components we consider 
\begin{align}
\mathbb{W}_\perp(\mathbf{P}=0,z_1,z_2) = W_0   \left[\begin{array}{cc} 
\frac{1}{2}|\alpha|^2 &0 \\
0 &1\end{array}\right] 
\end{align}
whereby
\begin{align}
\mathcal{D}_{2\txtpow{D}}^\perp = \frac{\sqrt{1 - 2 |\alpha|^2}}{1+|\alpha|^2}.
\end{align}

\section{Degree of coherence}
The degree of coherence of a 3D field is defined as $\mu(\mathbf{P},z_1,z_2) ={\tr[\mathbb{W}(\mathbf{P},z_1,z_2)] }/{\tr[\mathbb{W}(\mathbf{0},0,0)]}$.
Noting that $\text{tr}[\mathbb{V}_n] = (|\alpha|^2 + 1)L_{n0}$ taking the trace of Eq.~\eqref{eq:WSPPfinal} yields
\begin{align}
&\tr[\mathbb{W}(\mathbf{P},z_1,z_2)]=   \frac{(2\pi)^4\omega^2\mu_0^2}{|K_0|^2} (|\alpha|^2 + 1)  \int_0^\infty Q\,\{|\alpha|^2w_{11}(Q) +w_{33}(Q)\} L_{00}(P,Q,z_1,z_2) dQ.
\end{align}
	Letting $T_0 = {(2\pi)^4\omega^2\mu_0^2} (|\alpha|^2 + 1) / {|K_0|^2}$ and $\mathcal{W}(Q) = |\alpha|^2w_{11}(Q) +w_{33}(Q)$, we can split the integration domain such that 
	\begin{align}
	\tr[\mathbb{W}(\mathbf{P},z_1,z_2)]&=   T_0  \left[ \int_0^P Q\,\mathcal{W}(Q) {L}_{00}(P,Q,z_1,z_2)  d{Q}  +  \int_P^\infty Q\,\mathcal{W}(Q) {L}_{00}(P,Q,z_1,z_2)  d{Q}\right]. \label{eq:trW}
	\end{align}
	All values of $Q$ in the integration domain of the first term are such that $Q<P$ and the $L_{k0}$ factors are given by the first expression in Eq.~\eqref{eq:Lmn}, whereas the converse is true for the second integral in Eq.~\eqref{eq:trW}. Hence we can write
	\begin{align}
	&\tr[\mathbb{W}(\mathbf{P},z_1,z_2)]=   T_0  \sum_{p=1}^2  \Lambda_{00}(\kappa_p) \left[H^{(1)}_0(\kappa_p P) \int_0^P   Q\,\mathcal{W}(Q) {J}_{0}(\kappa_pQ)  d{Q}  +  J_0(\kappa_p P)\int_P^\infty  Q\,\mathcal{W}(Q) {H}^{(1)}_{0}(\kappa_p Q)  d{Q}\right] .
	\end{align}
	Noting that $\Lambda_{k0}(\kappa_2) = -(-1)^k \Gamma\Lambda_{00}(\kappa_1)$, where $\Gamma = (\kappa_2/\kappa_1)^2\exp[(\kappa_1+\kappa_2)\mathcal{Z}] $ we can further write
	\begin{align}
	&\tr[\mathbb{W}(\mathbf{P},z_1,z_2)]  =T_0 \Lambda_{00}(\kappa_1)  \sum_{p=1}^2   r_p \left[H^{(1)}_0(\kappa_p P) \int_0^P   Q\,\mathcal{W}(Q) {J}_{0}(\kappa_pQ)  d{Q} +  J_0(\kappa_p P)\int_P^\infty   Q\,\mathcal{W}(Q) {H}^{(1)}_{0}(\kappa_p Q)  d{Q}\right] 
	\end{align}
	where $r_1 = 1$ and $r_2 = - \Gamma$. Hence 
	\begin{align}
	\mu(\mathbf{P},z_1,z_2) =e^{-\kappa_1\mathcal{Z}} \frac{ \sum_{p=1}^2   r_p \left[H^{(1)}_0(\kappa_p P) \int_0^P  Q\,\mathcal{W}(Q) {J}_{0}(\kappa_pQ)  d{Q} +  J_0(\kappa_p P)\int_P^\infty   Q\,\mathcal{W}(Q) {H}^{(1)}_{0}(\kappa_p Q)  d{Q}\right]  }{   \sum_{p=1}^2   r_p \int_0^\infty  Q\,\mathcal{W}(Q) {H}^{(1)}_{0}(\kappa_p Q)  d{Q} }  \label{eq:DOCgeneral}.
	\end{align}
Eq.~\eqref{eq:DOCgeneral} gives a general expression for the DOC which is dependent on the surface and source correlation functions through $w_{11}=w_{22}$ and $w_{33}$, however, at this point we make a further approximation. Specifically, we split $\kappa_1 = k_{\txtpow{SPP}}$ into its real and imaginary parts viz. $\kappa_1 = \kappa_{1r} + i \kappa_{1i}$  and note that for typical SPPs $\kappa_{1i}\ll \kappa_{1r}$. We wish to approximate the integral terms in Eq.~\eqref{eq:DOCgeneral} and thus we expand the arguments of the Bessel and Hankel functions as $\kappa_{1r} Q + i \kappa_{1i} Q$. For $\kappa_{1i} Q \ll 1$ (or equivalently $ Q \ll 2 L$ where $L = 1 / (2\mbox{Im}[k_{\txtpow{SPP}}])$ is the SPP attenuation length) we can perform a series expansion about $\kappa_{1i} = 0$ to first order. Doing this yields an expression of the same form as Eq.~\eqref{eq:DOCgeneral} albeit with the replacements $J_0(\kappa_pQ)\rightarrow J_0(\kappa_{rp}Q)$ and similarly for the $H_0(\kappa_p Q)$ terms, where $\kappa_{r2} = - \kappa_{r1}$. Using the reflection formulae for the Bessel and Hankel functions \cite{Abramowitz1972} we have $J_0(\kappa_{2r} Q) = J_0(\kappa_{1r} Q)$, $H_0^{(1)}(\kappa_{2r}Q ) = - H_0^{(2)}(\kappa_{1r}Q )$ yielding
\begin{align}
\mu(\mathbf{P},z_1,z_2) \approx\frac{1}{2} e^{-\kappa_1\mathcal{Z}} &\left\{\Big[ H^{(1)}_0(\kappa_1 P) - H^{(1)}_0(\kappa_2 P) \Big] f_J^P(P) \right. \nonumber\\
&\quad\left.+ \Big[ J_0(\kappa_1 P) + J_0(\kappa_2 P)  \Big]f_J^\infty(P) + i \Big[ J_0(\kappa_1 P) - J_0(\kappa_2 P)  \Big]f_Y^\infty(P)\right\} \label{eq:DOClowloss}
\end{align}
where 
\begin{align}
f_J^P(P) &= \frac{\int_0^P  Q\,\mathcal{W}(Q) J_{0}(\kappa_{1r} Q)dQ}{\int_0^\infty  Q\,\mathcal{W}(Q) J_{0}(\kappa_{1r} Q)dQ} \label{eq:ffac1}\\
f_J^\infty(P) &= \frac{\int_P^\infty  Q\,\mathcal{W}(Q) J_0(\kappa_{1r} Q)dQ}{ \int_0^\infty  Q\,\mathcal{W}(Q) J_{0}(\kappa_{1r} Q)dQ} = 1-f_J^P(P)\label{eq:ffac2}\\
f_Y^\infty(P) &= \frac{ \int_P^\infty  Q\,\mathcal{W}(Q) Y_{0}(\kappa_{1r} Q)dQ}{ \int_0^\infty  Q\,\mathcal{W}(Q) J_{0}(\kappa_{1r} Q)dQ}\label{eq:ffac3}
\end{align}
Upon making a number of further approximations we can simplify Eq.~\eqref{eq:DOCgeneral} for two cases, namely that of (i) a loss-free metal such that $\epsilon_2$ is purely real and negative, and (ii) a narrow correlation function $w_{jk}$. We shall consider both cases now in turn.

\subsection{Loss-free case}
If we assume that the electric permittivity of the metal is purely real and negative then consequently $k_{\txtpow{SPP}}$ and $\alpha$ are  purely real (the dielectric is also assumed lossless) then $\kappa_1 = -\kappa_2$. Accordingly Eq.~\eqref{eq:DOClowloss} is exact. We can however simplify the terms in square brackets appearing in Eq.~\eqref{eq:DOClowloss} further by again recalling the reflection formulae for the Bessel and Hankel functions. Specifically we have 
\begin{align}
H^{(1)}_0(\kappa_1 P) - H^{(1)}_0(\kappa_2 P)  &=  H^{(1)}_0(\kappa_{1r} P) + H^{(2)}_0(\kappa_{1r}   P) = 2 J_0(\kappa_{1r} P) \\
J_0(\kappa_1 P) + J_0(\kappa_2 P) &=  J_0(\kappa_{1r} P) + J_0(\kappa_{1r} P) = 2 J_0(\kappa_{1r} P) \\
J_0(\kappa_1 P) - J_0(\kappa_2 P) &=  J_0(\kappa_{1r} P) - J_0(\kappa_{1r} P) =0
\end{align}
Substituting these results into Eq.~\eqref{eq:DOClowloss} and using Eq.~\eqref{eq:ffac2} it follows that the DOC reduces to the universal form
\begin{align}
\mu(\mathbf{P},z_1,z_2) =J_0(\kappa_1 P) \label{eq:DOCreal2}
\end{align}
irrespective of the form of the correlation matrix $\mathbbold{w}_J$ as has been previously shown for scalar 2D waves \cite{Ponomarenko2002}.

\subsection{Narrow correlation function}

Typically the correlation functions $w_{11}(Q)=w_{22}(Q)$ and $w_{33}(Q)$ fall-off to negligible values as $Q$ increases such that we can define a characteristic length $Q_0$ above which $w_{11}(Q) \approx 0$. When considering points $\mathbf{P}$ such that $P > Q_0$ we can note that the second term in both Eq.~\eqref{eq:DOCgeneral} is negligible such that 
\begin{align}
\!\!\!\mu(\mathbf{P},z_1,z_2) &=e^{-\kappa_1\mathcal{Z}} \frac{ \sum_{p=1}^2   r_p H^{(1)}_0(\kappa_p P) \int_0^P  Q\,\mathcal{W}(Q) {J}_{0}(\kappa_pQ)  d{Q}  }{   \sum_{p=1}^2   r_p \int_0^\infty  Q\,\mathcal{W}(Q) {H}^{(1)}_{0}(\kappa_p Q)  d{Q} }  \\
&\approx \frac{1}{2}e^{-\kappa_1\mathcal{Z}} \Big[ H^{(1)}_0(\kappa_1 P) - H^{(1)}_0(\kappa_2 P) \Big] f_J^P(P)\label{eq:DOCnarrow2}.
\end{align}
Since, however, for $ Q > Q_0$ we have assumed that the $w_{11}(Q) \approx w_{33}(Q) \approx 0$, the integration limits in Eq.~\eqref{eq:ffac1} can both be replaced by $\int_0^{Q_0} \cdots dQ$ without affecting the result. Accordingly $f_J^P = 1$ for all $P > Q_0$ and the DOC reduces to the form 
\begin{align}
\mu(\mathbf{P},z_1,z_2) &\approx \frac{1}{2}e^{-\kappa_1\mathcal{Z}} \left[ H^{(1)}_0(\kappa_1 P) - H^{(1)}_0(\kappa_2 P) \right].\label{eq:DOCnarrow}
\end{align}
It is worthwhile to note at this point that the DOC for points separated by a distance greater than the characteristic correlation length $Q_0$ is described by a form that is independent of the precise nature of the source and surface correlation functions. This statement is however only true for $P\gtrsim Q_0$, meaning when considering more closely spaced points the DOC between these positions exhibits non-universal behaviour i.e. the precise functional form of $\mu$ depends on $\mathbbold{w}_J$. Nevertheless, for very narrow correlation functions (i.e. $Q_0 \rightarrow 0$) the DOC is given by Eq.~\eqref{eq:DOCnarrow} for all $P \neq 0$. By construction we note that $\mu(\mathbf{0},0,0)=1$. 
Importantly, Eq.~\eqref{eq:DOCnarrow} does not hold in the large loss limit as can be seen by studying the limiting behaviour as $P\rightarrow 0$. In particular
\begin{align}
\lim_{P\rightarrow 0}  \frac{1}{2}\left[ H^{(1)}_0(\kappa_1 P) - H^{(1)}_0(\kappa_2 P) \right] = \frac{i}{\pi} \left[\log(\kappa_1) - \log(\kappa_2)\right] 
\end{align}
which does not equal unity for $k_{1i} >0$.